Three-dimensional Ising models- Critical Parameters using ε- convergence method

M V Vismaya[1] and M V Sangaranarayanan*

Department of Chemistry

Indian Institute of Technology Madras –Chennai 600036 India

1- vismayaviswanathan11@gmail.com; * sangara@iitm.ac.in

Abstract

We demonstrate the applicability of the ε- convergence algorithm in extracting the critical temperatures and critical exponents of three-dimensional Ising models. We analyze the low temperature magnetization as well as high temperature susceptibility series of simple cubic, body-centered cubic, face-centered cubic and diamond lattices, using *two different variables* for the inverse critical temperature. In the case of simple cubic lattices, the magnetization series was modified to deduce accurate values of the critical temperatures. The alternate variable for dimensionless inverse temperature suggested by Guttmann and Thompson has also been employed for the estimation of the critical parameters.

1. Introduction

The estimation of critical temperatures and exponents for the spontaneous magnetization in three-dimensional Ising models has been a frontier area of research in statistical physics [1]. In contrast to the Onsager's exact solution of two-dimensional Ising models in this context [2], the quantitative analysis of thermodynamic properties of three-dimensional systems has often resulted from extensive computer simulations of diverse genre' [3-9] as well as highly accurate series expansions [10-16].Employing effective field theory, Tambas [17] has provided a quantitative analytical expression for the spontaneous magnetization of face-centered cubic lattices that yield remarkable agreement with the simulation data. In an analogous manner, Kaya has reported magnetization equations for simple cubic [18] and body-centered cubic lattices [19].Viswanathan et al [20] have highlighted the pre-requisites for an acceptable solution of three-dimensional Ising models, apart from emphasizing the bottlenecks associated with simple transcription of the two-dimensional results to higher dimensions.

In an earlier article [21], we have proposed spontaneous magnetization equations for three-dimensional Ising models. Although the agreement with the simulation data was satisfactory, these equations do not directly yield the well-known low temperature expansions. The latter is a stringent requirement for the validity of any new expressions for the spontaneous magnetizations of simple cubic (s.c), body-centered cubic, (b.c.c), face-centered cubic (f.c.c) and diamond(d.) lattices.

As pointed out by Butera et al [22], despite the availability of a large number of series coefficients, the mathematical analysis 'remains arduous', due to the precise identification of the physical singularity. The exact values of critical exponents are often required for validating the universality and scaling hypothesis.

The series coefficients pertaining to diamond lattices follow a regular pattern in contrast to those for s.c, b.c.c and f.c.c lattices and among these, extensive studies have been made for s.c lattices using diverse approaches.

2. Methodology

The ε- convergence algorithm due to Shanks [23] and Wynn [24] is often employed for accelerating the convergence of a series or for obtaining the limit of a sequence. Its validity for obtaining the critical temperatures and exponents for a few Ising models has also been demonstrated [25]. In addition, the algorithm is also useful for predicting the densities of closest packing for virial equations of state of hard spheres and hard disks. [26]. It has been found to be particularly effective in obtaining the eigenvalues of anharmonic oscillators [27].

In this work, we employ the ε- convergence algorithm (EA) to estimate the critical temperatures and critical exponents for the series expansions pertaining to low temperature magnetization and high temperature susceptibility of the s.c, b.c.c, f.c.c and d. lattices. For both series expansions, more accurate results are obtained by incorporating the new low temperature variable suggested by Guttmann and Thompson [28]. The magnetization series for simple cubic lattices was rewritten in order to obtain satisfactory critical temperatures.

2. Methodology for estimating critical parameters

The ε- convergence algorithm involves construction of a two-dimensional array of elements, designated as ε-table. While the subscript k of $\varepsilon_k^j$ refers to the column, the superscript j indicates the systematic progression down the $j^{th}$ column. By an iterative procedure, the entire table is constructed in the following manner:

$$\varepsilon_{k+1}^j = \varepsilon_{k-1}^{j+1} + [\varepsilon_k^{j+1} - \varepsilon_k^j]^{-1}. \quad (1)$$

$$j = 0,1,\ldots; k = 0,1,\ldots,$$

where $\varepsilon_{-1}^j = 0$ for all $j$ values. The resulting Table is analogous to the difference table albeit with a nonlinear sequence transformation. The radius of convergence of the power series $\sum a_n x^n$ is then evaluated as limit n →∞ of the ratio $-\frac{a_n}{a_{n+1}}$ or $-\frac{a_{n+1}}{a_n}$.

**(A) Analysis of low temperature magnetization series**

**(i) Estimation of the critical temperature $x_c$ using the variable $x = e^{-4K}$.**

In the absence of exact solutions of three-dimensional Ising models, it is customary to study critical phenomena using series expansions. Furthermore, EA is more suitable when applied to the logarithmic derivative of a series, due to the conversion of an algebraic singularity into a simple pole [29]. Hence, we use $\frac{d}{dx}[\ln(M_0)]$, for deducing $x_c$. For low temperature series expansions, the preferred variable $(x)$ is $e^{-4K}$ where $K = \frac{J}{kT}$ and $x$ at the critical temperature $(T_c)$ is denoted as $x_c$. Hence

$$\frac{d}{dx}[\ln(M_0)] = \sum_{n=0}^{N} a_n x^n \quad (2)$$

with $x = e^{-4K}$. The known magnetization series expansions were employed for b.c.c, f.c.c and d. lattices [16]. However, for s.c lattices alone, the derivative of $\ln(M_0)$ led to erroneous results for $x_c$ while using EA. Since suitable transformations of a given series by redefining the variables may sometimes yield satisfactory convergence limits, several attempts were made to rewrite the magnetization series expansion. It was observed that if $x$ in equation (2) is replaced by $x \ln 2$, the derivative of series for $\frac{d}{dx}[\ln(M_0)]$ led to the correct estimate of $x_c$. The coefficients $\{a_n\}$ are shown in Table 1 for various lattices.

**Table 1: Coefficients $a_n$ for three-dimensional lattices (cf. Eqn 2)**

|   | $a_n$ | | | |
|---|---|---|---|---|
| **n** | **b.c.c** | **f.c.c** | **s.c** | **d.** |
| 0 | 0 | 0 | 0 | 0 |

| n | | | | |
|---|---|---|---|---|
| 1 | 0 | 0 | 0 | -4 |
| 2 | 0 | 0 | -1.99814791193 | -24 |
| 3 | -8 | 0 | 0 | -112 |
| 4 | 0 | 0 | -9.60016186543 | -480 |
| 5 | 0 | -12 | 7.98519015593 | -2128 |
| 6 | -112 | 0 | -48.4304803746 | -9408 |
| 7 | 128 | 0 | 71.6148286397 | -41792 |
| 8 | 0 | 0 | -254.846245721 | -185136 |
| 9 | -1680 | 0 | 485.392367970 | -815864 |
| 10 | 3872 | -264 | -1407.75952570 | -3593480 |
| 11 | -3368 | 288 | 3033.47814133 | -15810976 |
| 12 | -21216 | 0 | -8013.32290051 | -69510584 |
| 13 | 81200 | 0 | 18344.1331977 | -305621544 |
| 14 | -131280 | -720 | -46347.1752736 | |
| 15 | -171776 | -4032 | 109273.452793 | |
| 16 | 1377408 | 11424 | -270226.387563 | |
| 17 | -3326976 | -6996 | 646329.990945 | |
| 18 | 974928 | -3648 | -1581982.15998 | |
| 19 | 19059432 | -19680 | 3810379.40208 | |
| 20 | -67833136 | -23184 | -9280745.29781 | |
| 21 | 84324416 | 266904 | 22430563.0171 | |
| 22 | 189458176 | -404064 | -54506424.9469 | |
| 23 | -1176388384 | 46752 | 131961275.441 | |
| 24 | 2433544800 | -216600 | -320324452.195 | |
| 25 | 236151136 | 612924 | 776191434.561 | |
| 26 | | 3956256 | -1883260738.48 | |
| 27 | | -12891984 | 4565561047.96 | |
| 28 | | 9945144 | -11075406849.2 | |
| 29 | | -2120472 | 26857349367.6 | |
| 30 | | 21137784 | | |
| 31 | | 21780864 | | |
| 32 | | -280623024 | | |
| 33 | | 466770768 | | |
| 34 | | -211837080 | | |
| 35 | | 369977328 | | |
| 36 | | -692426436 | | |
| 37 | | -4023689280 | | |

In Pade' Approximants, the order of the Pade' Approximation plays a crucial role in dictating the accuracy of the desired parameters. In EA too, the choice of the coefficients is pivotal. In view of the large number of available coefficients shown in Table 1, it is not easy to decipher their effects. We have provided a few ranges of coefficients for each lattice, that led to the mean values of the critical temperatures and exponents. (Appendix)

A few typical ε- tables are shown in Tables 2,3,4 and 5 for b.c.c, f.c.c, s.c and d. lattices. The ratios $\left\{-\frac{a_n}{a_{n+1}}\right\}$ were employed for b.c.c (Table 2 using $a_8$ to $a_{19}$), s.c (Table 4 using $a_{18}$ to $a_{27}$) and d. (Table 5 using $a_1$ to $a_{12}$) lattices and the ratio $\left\{-\frac{a_{n+1}}{a_n}\right\}$ for f.c.c lattices ( Table 3 using $a_{25}$ to $a_{31}$)

### Table 2: ε-Table for b.c.c lattice

| $\varepsilon_0^j$ | $\varepsilon_1^j$ | $\varepsilon_2^j$ | $\varepsilon_3^j$ | $\varepsilon_4^j$ | $\varepsilon_5^j$ | $\varepsilon_6^j$ | $\varepsilon_7^j$ | $\varepsilon_8^j$ | $\varepsilon_9^j$ | $\varepsilon_{10}^j$ | $\varepsilon_{11}^j$ |
|---|---|---|---|---|---|---|---|---|---|---|---|
| 0 |  |  |  |  |  |  |  |  |  |  |  |
| 0 | 0 |  |  |  |  |  |  |  |  |  |  |
| 0 | 0.433884 | 2.304761 |  |  |  |  |  |  |  |  |  |
| 0 | 1.149643 | 1.397117 | -0.667868 |  |  |  |  |  |  |  |  |  |
| 0 | -0.158748 | -0.764297 | 0.686983 | 2.135205 |  |  |  |  |  |  |  |  |
| 0 | 0.261280 | 2.380788 | 0.159208 | -2.659042 | 0.478400 |  |  |  |  |  |  |  |
| 0 | 0.618525 | 2.799203 | 2.651253 | 2.782065 | 0.342994 | -10.044236 |  |  |  |  |  |
| 0 | -0.764251 | -0.723182 | 0.334626 | 2.367541 | 0.238848 | -6.819871 | 0.653132 |  |  |  |  |
| 0 | 0.124709 | 1.124909 | -0.223152 | -2.516006 | 0.129857 | -6.807523 | 81.225405 | -6.807460 |  |  |  |
| 0 | 0.414012 | 3.456590 | 0.553584 | 2.412346 | -0.020244 | -9.178118 | -0.291977 | -6.819790 | 0.127741 |  |  |
| 0 | 3.412535 | 0.333497 | 0.093816 | 1.281580 | -0.330772 | -0.807983 | 0.092274 | -6.621913 | 4.761647 | -6.603990 |  |
| 0 | -0.051151 | -0.288709 | 1.805353 | 0.917767 | -2.654848 | 0.851302 | 0.271896 | 4.983430 | 0.185394 | -6.840432 | 0.53228 |
| 0 |  |  |  |  |  |  |  |  |  |  |  |

### Table 3: ε-Table for f.c.c lattice

| $\varepsilon_0^j$ | $\varepsilon_1^j$ | $\varepsilon_2^j$ | $\varepsilon_3^j$ | $\varepsilon_4^j$ | $\varepsilon_5^j$ | $\varepsilon_6^j$ |
|---|---|---|---|---|---|---|
| 0 |  |  |  |  |  |  |
| 0 | -6.454725 |  |  |  |  |  |
| 0 | 3.258632 | 0.102951 |  |  |  |  |
| 0 | 0.771420 | -0.402056 | 1.278464 |  |  |  |
| 0 | 0.213216 | -1.791459 | 0.051687 | -1.217200 |  |  |
| 0 | 9.968433 | 0.102509 | 0.741208 | -0.341178 | 1.193210 |  |
| 0 | -1.030423 | -0.090918 | 4.798545 | 0.348976 | 2.190159 | 0.66188 |
| 0 |  |  |  |  |  |  |

### Table 4: ε- Table for s.c lattice

| $\varepsilon_0^j$ | $\varepsilon_1^j$ | $\varepsilon_2^j$ | $\varepsilon_3^j$ | $\varepsilon_4^j$ | $\varepsilon_5^j$ | $\varepsilon_6^j$ | $\varepsilon_7^j$ | $\varepsilon_8^j$ | $\varepsilon_9^j$ |
|---|---|---|---|---|---|---|---|---|---|
| 0 |  |  |  |  |  |  |  |  |  |
| 0 | 0.415177 |  |  |  |  |  |  |  |  |
| 0 | 0.410568 | -216.97300 |  |  |  |  |  |  |  |
| 0 | 0.413755 | 313.84500 | 0.412452 |  |  |  |  |  |  |
| 0 | 0.411521 | -447.79800 | 0.412442 | -94472.70000 |  |  |  |  |  |
| 0 | 0.413049 | 654.74600 | 0.412428 | 76362.80000 | 0.412497 |  |  |  |  |
| 0 | 0.411961 | -919.55100 | 0.412413 | -66454.10000 | 0.412529 | -45635.30000 |  |  |  |
| 0 | 0.412688 | 1376.76000 | 0.412397 | -60413.20000 | 0.412579 | -46337.50000 | 0.411105 |  |  |
| 0 | 0.412153 | -1870.16000 | 0.412380 | -57080.40000 | 0.412697 | -51918.40000 | 0.412400 | -45565.10000 |  |
| 0 | 0.412493 | 2941.82000 | 0.412361 | -54795.10000 | 0.412817 | -48774.90000 | 0.413015 | -50292.40000 | 0.41218 |
| 0 |  |  |  |  |  |  |  |  |  |

Table 5: ε- Table for d. lattice

| $\varepsilon_0^j$ | $\varepsilon_1^j$ | $\varepsilon_2^j$ | $\varepsilon_3^j$ | $\varepsilon_4^j$ | $\varepsilon_5^j$ | $\varepsilon_6^j$ | $\varepsilon_7^j$ | $\varepsilon_8^j$ | $\varepsilon_9^j$ | $\varepsilon_{10}^j$ | $\varepsilon_{11}^j$ |
|---|---|---|---|---|---|---|---|---|---|---|---|
| 0 | | | | | | | | | | | |
| | 0.166666 | | | | | | | | | | |
| 0 | | 21.000000 | | | | | | | | | |
| | 0.214285 | | 0.246031 | | | | | | | | |
| 0 | | 52.500000 | | -2.394136 | | | | | | | |
| | 0.233333 | | 0.227814 | | 0.226434 | | | | | | |
| 0 | | -128.709677 | | -727.101167 | | -1194.318387 | | | | | |
| | 0.225563 | | 0.226143 | | 0.224294 | | 0.225591 | | | | |
| 0 | | 1596.000000 | | -1267.918474 | | -423.116018 | | 13.818701 | | | |
| | 0.226190 | | 0.225794 | | 0.225478 | | 0.227880 | | 0.226175 | | |
| 0 | | -929.694915 | | -4430.747689 | | -6.733747 | | -572.708275 | | 2668.241982 | |
| | 0.225114 | | 0.225508 | | 0.225704 | | 0.226113 | | 0.226483 | | 0.22818 |
| 0 | | 1607.972547 | | 684.477324 | | 2439.832038 | | 2126.609096 | | 3255.116438 | |
| | 0.225736 | | 0.224426 | | 0.226274 | | 0.222920 | | 0.227369 | | |
| 0 | | 845.010471 | | 1225.596165 | | 2141.641760 | | 2351.368378 | | | |
| | 0.226920 | | 0.227053 | | 0.227365 | | 0.227688 | | | | |
| 0 | | 8339.536756 | | 4429.261530 | | 5238.577051 | | | | | |
| | 0.227040 | | 0.226797 | | 0.228601 | | | | | | |
| 0 | | 4210.998799 | | 4983.741710 | | | | | | | |
| | 0.227277 | | 0.228091 | | | | | | | | |
| 0 | | 5438.894211 | | | | | | | | | |
| | 0.227461 | | | | | | | | | | |
| 0 | | | | | | | | | | | |

**(ii) Estimation of the critical temperature $x_c$ using the variable $x_1 = 1 - \tanh K$.**

As pointed out by Guttmann and Thompson, [28], a more natural variable is $1 - \tanh K$ for the analysis of low temperature expansions. We estimate $x_c$ by replacing $x$ of equation (3) with $x_1 = [1 - \tanh K]$. Hence

$$\frac{d\{ln(M_0)\}}{dx_1} = \sum_{n=0}^{N} b_n x_1^n \qquad (3)$$

The coefficients $\{b_n\}$ are shown in Table 6.

Table 6: Coefficients $b_n$ for three-dimensional lattices (cf.Eqn 3)

| | $b_n$ | | | |
|---|---|---|---|---|
| n | b.c.c | f.c.c | s.c | d. |
| 0 | 0 | 0 | 0 | 0 |
| 1 | 0 | 0 | 0 | 0 |
| 2 | 0 | 0 | 0 | 0 |
| 3 | 0 | 0 | 0 | -0.5 |
| 4 | 0 | 0 | 0 | -1.25 |
| 5 | 0 | 0 | -0.1875 | -2.625 |
| 6 | 0 | 0 | -0.65625 | -4.8125 |
| 7 | -0.0625 | 0 | -1.3125 | -8.3125 |
| 8 | -0.28125 | 0 | -1.96875 | -13.78125 |
| 9 | -0.703125 | 0 | -2.578125 | -22.265625 |
| 10 | -1.2890625 | 0 | -3.3515625 | -35.3203125 |
| 11 | -1.93359375 | -0.005859 | -4.60546875 | -55.37304688 |
| 12 | -2.513671875 | -0.038086 | -6.47460938 | -86.21386719 |
| 13 | -2.946289063 | -0.133301 | -8.80810547 | -133.7827148 |
| 14 | -3.244628906 | -0.333252 | -11.4074707 | -207.3425293 |
| 15 | -3.548339844 | -0.666504 | -14.3964844 | -321.3032227 |
| 16 | -4.096435547 | -1.133057 | -18.328125 | -498.0510254 |
| 17 | -5.136108398 | -1.699585 | -23.8748474 | -772.3579102 |
| 18 | -6.801452637 | -2.30658 | -31.402359 | -1198.231567 |
| 19 | -9.022521973 | -2.883224 | -40.9073639 | -1859.537743 |
| 20 | -11.52218628 | -3.363762 | -52.4633102 | -2886.443661 |
| 21 | -13.92657471 | -3.700264 | -66.7744761 | -4480.856676 |
| 22 | -15.96305847 | -3.869774 | -85.2735744 | -6955.780576 |
| 23 | -17.66568756 | -3.876978 | -109.648038 | -10796.22987 |

| | | | |
|---|---|---|---|
| 24 | -19.47984695 | -3.755307 | -141.319835 | -16753.54951 |
| 25 | -22.18110371 | -3.568697 | -181.53928 | -25991.49245 |
| 26 | -26.60107183 | -3.413508 | -232.169502 | -40312.46245 |
| 27 | | -3.417012 | | -62507.88166 |
| 28 | | -3.727509 | | -96900.71459 |
| 29 | | -4.492857 | | -150185.4585 |
| 30 | | -5.828599 | | -232728.6567 |
| 31 | | -7.782055 | | |
| 32 | | -10.30235 | | |
| 33 | | -13.2267 | | |
| 34 | | -16.2903 | | |
| 35 | | -19.16173 | | |
| 36 | | -21.50001 | | |
| 37 | | -23.02487 | | |
| 38 | | -23.58874 | | |
| 39 | | -23.23798 | | |
| 40 | | -22.25101 | | |

The $b_n$ coefficients were utilized to calculate the ratios $\left\{\frac{b_n}{b_{n+1}}\right\}$ in the ε-table. These are displayed in Tables 7, 8, 9, and 10, respectively, for the lattices b.c.c ($b_{17}$ to $b_{26}$), f.c.c ($b_{23}$ to $b_{32}$), s.c ($b_{11}$ to $b_{22}$), and d. ($b_4$ to $b_{13}$).

**Table 7: ε- Table for b.c.c lattice**

| $\varepsilon_0^j$ | $\varepsilon_1^j$ | $\varepsilon_2^j$ | $\varepsilon_3^j$ | $\varepsilon_4^j$ | $\varepsilon_5^j$ | $\varepsilon_6^j$ | $\varepsilon_7^j$ | $\varepsilon_8^j$ | $\varepsilon_9^j$ |
|---|---|---|---|---|---|---|---|---|---|
| 0 | | | | | | | | | |
| | 0.755148 | | | | | | | | |
| 0 | | -758.611065 | | | | | | | |
| | 0.753830 | | 0.755091 | | | | | | |
| 0 | | 34.216249 | | 16.956887 | | | | | |
| | 0.783056 | | 0.697152 | | 0.889137 | | | | |
| 0 | | 22.575367 | | 22.165614 | | 5.619218 | | | |
| | 0.827352 | | -1.743345 | | 0.828701 | | 0.739454 | | |
| 0 | | 22.186367 | | 22.554409 | | -5.585571 | | 8.068044 | |
| | 0.872425 | | 0.973733 | | 0.793165 | | 0.812694 | | 0.84523 |
| 0 | | 32.057228 | | 17.016344 | | 45.618352 | | 38.791292 | |
| | 0.903619 | | 0.907248 | | 0.828127 | | 0.666218 | | |
| 0 | | 307.644126 | | 4.377386 | | 39.442034 | | | |
| | 0.906869 | | 0.903950 | | 0.856646 | | | | |
| 0 | | -34.901788 | | -16.762357 | | | | | |
| | 0.878218 | | 0.959079 | | | | | | |
| 0 | | -22.534891 | | | | | | | |
| | 0.833842 | | | | | | | | |
| 0 | | | | | | | | | |

**Table 8: ε- Table for f.c.c lattice**

| $\varepsilon_0^j$ | $\varepsilon_1^j$ | $\varepsilon_2^j$ | $\varepsilon_3^j$ | $\varepsilon_4^j$ | $\varepsilon_5^j$ | $\varepsilon_6^j$ | $\varepsilon_7^j$ | $\varepsilon_8^j$ | $\varepsilon_9^j$ |
|---|---|---|---|---|---|---|---|---|---|
| 0 | | | | | | | | | |
| | 1.032399 | | | | | | | | |
| 0 | | 50.273527 | | | | | | | |
| | 1.052290 | | 1.047207 | | | | | | |
| 0 | | -146.463407 | | 13.324251 | | | | | |
| | 1.045463 | | 1.053466 | | 0.989965 | | | | |
| 0 | | -21.510576 | | -2.423664 | | 225.706506 | | | |
| | 0.998974 | | 1.105858 | | 0.994349 | | 0.990322 | | |
| 0 | | -12.154612 | | -11.391555 | | -22.631335 | | -21.819072 | |
| | 0.916701 | | 2.416376 | | 0.905379 | | 2.221451 | | 0.89755 |
| 0 | | -11.48780 | | -12.053370 | | -21.871498 | | -22.574421 | |
| | 0.829652 | | 0.648245 | | 0.803527 | | 0.798820 | | |
| 0 | | -17.000262 | | -5.613473 | | -234.326044 | | | |
| | 0.770829 | | 0.736066 | | 0.799154 | | | | |
| 0 | | -45.766068 | | 10.237248 | | | | | |
| | 0.748979 | | 0.753922 | | | | | | |
| 0 | | 156.546876 | | | | | | | |
| | 0.755367 | | | | | | | | |
| 0 | | | | | | | | | |

## Table 9: ε- Table for s.c lattice

| $\varepsilon_0^j$ | $\varepsilon_1^j$ | $\varepsilon_2^j$ | $\varepsilon_3^j$ | $\varepsilon_4^j$ | $\varepsilon_5^j$ | $\varepsilon_6^j$ | $\varepsilon_7^j$ | $\varepsilon_8^j$ | $\varepsilon_9^j$ | $\varepsilon_{10}^j$ | $\varepsilon_{11}^j$ |
|---|---|---|---|---|---|---|---|---|---|---|---|
| 0 | | | | | | | | | | | |
| | 0.711312 | | | | | | | | | | |
| 0 | | 42.084380 | | | | | | | | | |
| | 0.735074 | | 0.668856 | | | | | | | | |
| 0 | | 26.982670 | | 33.744340 | | | | | | | |
| | 0.772134 | | 0.816748 | | 0.761899 | | | | | | |
| 0 | | 49.397150 | | 15.512690 | | 153.621100 | | | | | |
| | 0.792378 | | 0.787236 | | 0.769139 | | 0.775414 | | | | |
| 0 | | -145.075000 | | -39.745300 | | 312.988600 | | -80.705270 | | | |
| | 0.785485 | | 0.796730 | | 0.771974 | | 0.772874 | | 0.774267 | | |
| 0 | | -56.145500 | | -80.139660 | | 1424.392000 | | 637.223200 | | 597.592300 | |
| | 0.767675 | | 0.755054 | | 0.772639 | | 0.771604 | | 0.749034 | | 0.78229 |
| 0 | | -135.378100 | | -23.74720 | | 458.465800 | | 592.915700 | | 627.660900 | |
| | 0.760288 | | 0.763974 | | 0.774715 | | 0.779042 | | 0.777815 | | |
| 0 | | 135.919700 | | 69.826540 | | 689.592400 | | -222.541300 | | | |
| | 0.767645 | | 0.748844 | | 0.776328 | | 0.777945 | | | | |
| 0 | | 82.732330 | | 106.210500 | | 1308.103000 | | | | | |
| | 0.779732 | | 0.791437 | | 0.777160 | | | | | | |
| 0 | | 168.171800 | | 36.163470 | | | | | | | |
| | 0.785679 | | 0.783861 | | | | | | | | |
| 0 | | -382.068800 | | | | | | | | | |
| | 0.783061 | | | | | | | | | | |
| 0 | | | | | | | | | | | |

## Table 10: ε- Table for d. lattice

| $\varepsilon_0^j$ | $\varepsilon_1^j$ | $\varepsilon_2^j$ | $\varepsilon_3^j$ | $\varepsilon_4^j$ | $\varepsilon_5^j$ | $\varepsilon_6^j$ | $\varepsilon_7^j$ | $\varepsilon_8^j$ | $\varepsilon_9^j$ |
|---|---|---|---|---|---|---|---|---|---|
| 0 | | | | | | | | | |
| | 0.476190 | | | | | | | | |
| 0 | | 14.437500 | | | | | | | |
| | 0.545454 | | 0.610306 | | | | | | |
| 0 | | 29.857150 | | 47.645650 | | | | | |
| | 0.578947 | | 0.666522 | | 0.650254 | | | | |
| 0 | | 41.275850 | | -13.821440 | | 212.387200 | | | |
| | 0.603174 | | 0.648373 | | 0.654674 | | 0.654828 | | |
| 0 | | 63.400420 | | 144.869800 | | 6723.848000 | | -56.537660 | |
| | 0.618947 | | 0.660647 | | 0.65486 | | 0.654680 | | 0.64698 |
| 0 | | 87.380990 | | -26.921860 | | -129.834600 | | -186.492800 | |
| | 0.630391 | | 0.651899 | | 0.645109 | | 0.637031 | | |
| 0 | | 133.876200 | | -174.212800 | | -253.618000 | | | |
| | 0.637861 | | 0.648653 | | 0.632516 | | | | |
| 0 | | 226.535800 | | -236.181800 | | | | | |
| | 0.642275 | | 0.646492 | | | | | | |
| 0 | | 463.686400 | | | | | | | |
| | 0.644432 | | | | | | | | |
| 0 | | | | | | | | | |

**(iii) Estimation of $\beta$ using the variable $x$**

In order to deduce the critical exponent of the function $(x) \sim \left(1 - \frac{x}{x_c}\right)^\beta$, we form the new series

$$ln[f(x)] = \sum_{n=0}^{N} c_n \left(\frac{x}{x_c}\right)^n$$

and obtain the asymptotic limit of the sequence using $\{-(n+1)a_n\}$ or $\{-na_n\}$. The limit then yields $-\beta$ as the exponent. Employing $x_c = e^{-4K_c} = 0.53286, 0.66480, 0.41206$ and $0.22791$ for b.c.c, f.c.c, s.c and d. lattices[6,7], the new series for $ln(M_0)$ is written as

$$ln(M_0) = \sum_{n=0}^{N} c_n X^n, \text{ where } X = \frac{x}{x_c} \qquad (4)$$

The coefficients $\{c_n\}$ are shown in Table 11 for the b.c.c, f.c.c, s.c and d. lattices.

**Table 11: Coefficients $c_n$ for three-dimensional lattices (cf.Eqn 4)**

| n | $c_n$ b.c.c | f.c.c | s.c | d. |
|---|---|---|---|---|
| 0 | 0 | 0 | 0 | 0 |
| 1 | 0 | 0 | 0 | 0 |
| 2 | 0 | 0 | 0 | -0.10387024 |
| 3 | 0 | 0 | -0.13992443 | -0.094685116 |
| 4 | -0.1612543 | 0 | 0 | -0.075523192 |
| 5 | 0 | 0 | -0.14254560 | -0.059009796 |
| 6 | 0 | -0.17266313 | 0.058736535 | -0.049682545 |
| 7 | -0.1951921 | 0 | -0.18151984 | -0.042905533 |
| 8 | 0.1040118 | 0 | 0.13961928 | -0.038005616 |
| 9 | 0 | 0 | -0.26254103 | -0.034105432 |
| 10 | -0.3101076 | 0 | 0.26753676 | -0.030826452 |
| 11 | 0.3462311 | -0.26906396 | -0.41932944 | -0.028129298 |
| 12 | -0.1471074 | 0.17887533 | 0.49238888 | -0.025854989 |
| 13 | -0.4558103 | 0 | -0.71375137 | -0.023911404 |
| 14 | 0.8632018 | 0 | 0.90193632 | -0.022247643 |
| 15 | -0.6940839 | -0.10511514 | -1.2643514 | |
| 16 | -0.4536986 | -0.36687617 | 1.6613438 | |
| 17 | 1.824563 | 0.65040394 | -2.2986492 | |
| 18 | -2.217904 | -0.25008412 | 3.0867732 | |
| 19 | 0.3280989 | -0.082130599 | -4.2550038 | |
| 20 | 3.247029 | -0.27982969 | 5.7878764 | |
| 21 | -5.864745 | -0.20871935 | -7.9813000 | |
| 22 | 3.708322 | 1.5248284 | 10.946030 | |
| 23 | 4.246710 | -1.4679316 | -15.124744 | |
| 24 | -13.46564 | 0.10821000 | 20.860874 | |
| 25 | 14.24974 | -0.31995715 | | |
| 26 | | 0.57876410 | | |
| 27 | | 2.3915742 | | |
| 28 | | -4.9959709 | | |

The sequence $\{(n + 1)c_n\}$ was calculated for the lattices b.c.c ($c_{16}$ to $c_{22}$), f.c.c ($c_{15}$ to $c_{23}$), s.c ($c_{17}$ to $c_{23}$), and d. ($c_8$ to $c_{12}$), as the initial entries in the ε- table and are shown in Tables 12, 13, 14, and 15, respectively.

**Table 12: ε-Table for b.c.c lattice**

| $\varepsilon_0^j$ | $\varepsilon_1^j$ | $\varepsilon_2^j$ | $\varepsilon_3^j$ | $\varepsilon_4^j$ | $\varepsilon_5^j$ | $\varepsilon_6^j$ | $\varepsilon_7^j$ |
|---|---|---|---|---|---|---|---|
| 0 | | | | | | | |
| | -7.71287 | | | | | | |
| 0 | | 0.02465 | | | | | |
| | 32.84214 | | 6.52243 | | | | |
| 0 | | -0.01333 | | -0.06558 | | | |
| | -42.14017 | | -12.61504 | | -0.33447 | | |
| 0 | | 0.02053 | | 0.01583 | | 224.40917 | |
| | 6.56197 | | -225.67441 | | -0.33001 | | -0.32778 |
| 0 | | 0.01622 | | 0.02027 | | 673.08774 | |
| | 68.18760 | | 21.23418 | | -0.32852 | | |
| 0 | | -0.00507 | | -0.02609 | | | |
| | -129.02438 | | -26.32015 | | | | |
| 0 | | 0.00466 | | | | | |
| | 85.29141 | | | | | | |
| 0 | | | | | | | |

**Table 13: ε- Table for f.c.c lattice**

| $\varepsilon_0^j$ | $\varepsilon_1^j$ | $\varepsilon_2^j$ | $\varepsilon_3^j$ | $\varepsilon_4^j$ | $\varepsilon_5^j$ | $\varepsilon_6^j$ | $\varepsilon_7^j$ | $\varepsilon_8^j$ | $\varepsilon_9^j$ |
|---|---|---|---|---|---|---|---|---|---|
| 0 |  |  |  |  |  |  |  |  |  |
|  | -1.68184 |  |  |  |  |  |  |  |  |
| 0 |  | -0.21953 |  |  |  |  |  |  |  |
|  | -6.23689 |  | -2.60402 |  |  |  |  |  |  |
| 0 |  | -0.05572 |  | 0.23035 |  |  |  |  |  |
|  | 11.70727 |  | 3.12254 |  | 1.04464 |  |  |  |  |
| 0 |  | -0.06075 |  | -0.25090 |  | -0.37218 |  |  |  |
|  | -4.75159 |  | -2.13657 |  | -7.20055 |  | -5.30125 |  |  |
| 0 |  | 0.32164 |  | -0.44837 |  | 0.15432 |  | -0.35694 |  |
|  | -1.64261 |  | -3.43523 |  | -5.54135 |  | -7.25715 |  | -0.32646 |
| 0 |  | -0.23619 |  | -0.92318 |  | -0.42849 |  | -0.21266 |  |
|  | -5.87642 |  | -4.89085 |  | -3.51988 |  | -2.62389 |  |  |
| 0 |  | 0.77845 |  | -0.19377 |  | 0.68759 |  |  |  |
|  | -4.59182 |  | -5.91942 |  | -2.38528 |  |  |  |  |
| 0 |  | 0.02521 |  | 0.08917 |  |  |  |  |  |
|  | 35.07105 |  | 9.71412 |  |  |  |  |  |  |
| 0 |  | -0.01422 |  |  |  |  |  |  |  |
|  | -35.2303 |  |  |  |  |  |  |  |  |
| 0 |  |  |  |  |  |  |  |  |  |

**Table 14: ε- Table for s.c lattice**

| $\varepsilon_0^j$ | $\varepsilon_1^j$ | $\varepsilon_2^j$ | $\varepsilon_3^j$ | $\varepsilon_4^j$ | $\varepsilon_5^j$ | $\varepsilon_6^j$ | $\varepsilon_7^j$ |
|---|---|---|---|---|---|---|---|
| 0 |  |  |  |  |  |  |  |
|  | -41.37569 |  |  |  |  |  |  |
| 0 |  | 0.00999 |  |  |  |  |  |
|  | 58.64869 |  | -0.33393 |  |  |  |  |
| 0 |  | -0.00695 |  | 101.10450 |  |  |  |
|  | -85.10008 |  | -0.32404 |  | -0.32947 |  |  |
| 0 |  | 0.00483 |  | -83.00466 |  | 2300.18600 |  |
|  | 121.54540 |  | -0.33609 |  | -0.32905 |  | -0.32629 |
| 0 |  | -0.00336 |  | 59.14465 |  | 2661.67600 |  |
|  | -175.58860 |  | -0.31918 |  | -0.32867 |  |  |
| 0 |  | 0.00234 |  | -46.25564 |  |  |  |
|  | 251.75870 |  | -0.340805 |  |  |  |  |
| 0 |  | -0.00162 |  |  |  |  |  |
|  | -362.99390 |  |  |  |  |  |  |
| 0 |  |  |  |  |  |  |  |

**Table 15: ε- Table for d. lattice**

| $\varepsilon_0^j$ | $\varepsilon_1^j$ | $\varepsilon_2^j$ | $\varepsilon_3^j$ | $\varepsilon_4^j$ | $\varepsilon_5^j$ |
|---|---|---|---|---|---|
| 0 |  |  |  |  |  |
|  | -0.34205 |  |  |  |  |
| 0 |  | 1003.78731 |  |  |  |
|  | -0.34105 |  | -0.34307 |  |  |
| 0 |  | 509.33263 |  | 599.30335 |  |
|  | -0.33909 |  | -0.33196 |  | -0.32311 |
| 0 |  | 649.60586 |  | 712.29838 |  |
|  | -0.33755 |  | -0.31601 |  |  |
| 0 |  | 696.03022 |  |  |  |
|  | -0.33611 |  |  |  |  |
| 0 |  |  |  |  |  |

### (iv) Estimation of the critical exponent $\beta$ using the variable $x_1$

The methodology is identical to that outlined earlier and the new series is given by

$$\ln[f(x_1)] = \sum_{n=0}^{N} d_n \left(\frac{x_1}{(x_1)_c}\right)^n$$

The asymptotic limit from the sequence $\{-n d_n\}$, then yields $-\beta$ as the exponent.

Employing $(x_1)_c = 1 - \tanh K_c$ =0.843915, 0.898287, 0.781909 and 0.64625 for b.c.c, f.c.c, s.c and d. lattice respectively, the new series for $\ln(M_0)$ may be written as

$$\ln(M_0) = \sum_{n=0}^{N} d_n X_1^n, \text{ where } X_1 = \frac{x_1}{(x_1)_c} \tag{5}$$

The series coefficients are shown in Table 16. In this case, the sequence $\{-n d_n\}$ with various ranges of $n$ was formed as the initial entries in the ε-table.

**Table 16: Coefficients $d_n$ for three-dimensional lattices (cf. Eqn 5)**

| | $d_n$ | | | |
|---|---|---|---|---|
| n | b.c.c | f.c.c | s.c | d. |
| 0 | 0 | 0 | 0 | 0 |
| 1 | 0 | 0 | 0 | 0 |
| 2 | 0 | 0 | 0 | 0 |
| 3 | 0 | 0 | 0 | 0 |
| 4 | 0 | 0 | 0 | -0.021803230 |
| 5 | 0 | 0 | 0 | -0.028180820 |
| 6 | 0 | 0 | -0.0071415 | -0.031870910 |
| 7 | 0 | 0 | -0.016752 | -0.032366200 |
| 8 | -0.0020099320 | 0 | -0.022923 | -0.031612840 |
| 9 | -0.0067848490 | 0 | -0.023898 | -0.030107213 |
| 10 | -0.012883140 | 0 | -0.022023 | -0.028291898 |
| 11 | -0.018120460 | 0 | -0.020351 | -0.026367066 |
| 12 | -0.021026690 | -0.00013479000 | -0.020043501 | -0.024487808 |
| 13 | -0.021293700 | -0.00072648000 | -0.020337634 | -0.022744122 |
| 14 | -0.019558340 | -0.0021209000 | -0.020089019 | -0.021179223 |
| 15 | -0.016965140 | -0.0044454000 | -0.018986674 | -0.019798762 |
| 16 | -0.014678700 | -0.0074874000 | -0.017564349 | -0.018588259 |
| 17 | -0.013459807 | -0.010761000 | -0.01645608 | -0.017525533 |
| 18 | -0.013450611 | -0.013695000 | -0.01583039 | -0.016588044 |
| 19 | -0.014240571 | -0.015817000 | -0.015423394 | -0.015755739 |
| 20 | -0.015145234 | -0.016872000 | -0.014924887 | -0.015011689 |
| 21 | -0.015545070 | -0.016840000 | -0.014253379 | -0.014341716 |
| 22 | -0.015135531 | -0.015884000 | -0.013541016 | -0.013734037 |
| 23 | -0.014004352 | -0.014273000 | -0.012933059 | -0.013178946 |
| 24 | -0.012534103 | -0.012310009 | -0.012461414 | -0.012668531 |
| 25 | -0.011197429 | -0.010282098 | -0.012055344 | -0.012196491 |
| 26 | -0.010346212 | -0.0084399498 | -0.011643185 | -0.011757871 |
| 27 | -0.010083374 | -0.0069832400 | -0.011211696 | -0.011348779 |
| 28 | -0.010258840 | -0.0060551878 | -0.010795808 | |
| 29 | -0.010572458 | -0.0057289182 | -0.010427958 | |
| 30 | -0.010724112 | -0.0059961244 | -0.010108435 | |
| 31 | -0.010540251 | -0.0067621886 | -0.00981483 | |
| 32 | -0.010026967 | -0.0078567782 | -0.009527848 | |
| 33 | -0.0093382160 | -0.0090601250 | -0.009245892 | |
| 34 | -0.0086858336 | -0.010141511 | -0.008978423 | |
| 35 | -0.0082386494 | -0.010899982 | -0.008732528 | |
| 36 | -0.0080556616 | -0.011196888 | -0.008506196 | |
| 37 | -0.0080768499 | -0.010980472 | -0.008292189 | |
| 38 | -0.0081669768 | -0.010285061 | -0.008085479 | |
| 39 | -0.0081854685 | -0.0092224862 | -0.007885842 | |
| 40 | -0.0080482707 | -0.0079572799 | -0.007695873 | |
| 41 | -0.0077559957 | -0.0066773806 | -0.007516818 | |
| 42 | -0.0073810769 | -0.0055636090 | -0.007347367 | |
| 43 | -0.0070254438 | -0.0047617251 | -0.007185361 | |
| 44 | -0.0067714402 | -0.0043617619 | -0.007029116 | |
| 45 | -0.0066483431 | -0.0043862881 | -0.006878699 | |
| 46 | -0.0066268585 | -0.0047901028 | -0.006734709 | |
| 47 | -0.0066400487 | -0.0054695215 | -0.006597038 | |
| 48 | -0.0066177745 | -0.0062809472 | -0.00646517 | |
| 49 | -0.0065177168 | -0.0070659497 | -0.006338289 | |
| 50 | | -0.0076770549 | -0.006215795 | |
| 51 | | -0.0080017969 | -0.006097715 | |
| 52 | | -0.0079797932 | -0.005984163 | |
| 53 | | -0.0076109994 | -0.005874784 | |
| 54 | | -0.0069527915 | -0.005769447 | |
| 55 | | -0.0061089522 | -0.005667715 | |
| 56 | | -0.0052099971 | -0.005569323 | |
| 57 | | -0.0043910888 | | |
| 58 | | -0.0037692411 | | |
| 59 | | -0.0034248999 | | |
| 60 | | -0.0033901895 | | |
| 61 | | -0.0036453159 | | |
| 62 | | -0.0041236045 | | |
| 63 | | -0.0047246099 | | |
| 64 | | -0.0053315534 | | |
| 65 | | -0.0058311261 | | |
| 66 | | -0.0061325356 | | |
| 67 | | -0.0061811855 | | |
| 68 | | -0.0059665753 | | |
| 69 | | -0.0055230027 | | |
| 70 | | -0.0049216711 | | |
| 71 | | -0.0042578523 | | |
| 72 | | -0.0036343461 | | |
| 73 | | -0.0031441158 | | |
| 74 | | -0.0028548462 | | |

| | | | | -0.0027990707 | | | |
|---|---|---|---|---|---|---|---|
| 75 | | | | -0.0027990707 | | | |
| 76 | | | | -0.0029693172 | | | |
| 77 | | | | -0.0033211169 | | | |
| 78 | | | | -0.0037812946 | | | |
| 79 | | | | -0.0042612904 | | | |

For the lattices b.c.c ($d_{41}$ to $d_{49}$), f.c.c ($d_{71}$ to $d_{79}$), s.c ($d_{30}$ to $d_{38}$), and d. ($d_4$ to $d_8$), the entries of the ε- Table are provided in Tables 17, 18, 19, and 20 respectively.

Table 17: ε- Table for b.c.c lattice

| $\varepsilon_0^j$ | $\varepsilon_1^j$ | $\varepsilon_2^j$ | $\varepsilon_3^j$ | $\varepsilon_4^j$ | $\varepsilon_5^j$ | $\varepsilon_6^j$ | $\varepsilon_7^j$ | $\varepsilon_8^j$ | $\varepsilon_9^j$ |
|---|---|---|---|---|---|---|---|---|---|
| 0 | | | | | | | | | |
| 0 | -0.31800 | | | | | | | | |
| 0 | -0.31001 | 125.14000 | | | | | | | |
| 0 | -0.30209 | 126.40000 | 0.48387 | | | | | | |
| 0 | -0.29794 | 240.91000 | -0.29336 | 125.12000 | | | | | |
| 0 | -0.29918 | -811.28000 | -0.29889 | 60.15600 | -0.30875 | | | | |
| 0 | -0.30484 | -176.69000 | -0.29760 | -38.27600 | -0.30905 | -3295.30000 | | | |
| 0 | -0.31208 | -137.99000 | -0.27900 | -122.94000 | -0.30941 | -2823.30000 | -0.30693 | | |
| 0 | -0.31765 | -179.50000 | -0.33617 | -155.48000 | -0.30972 | -3324.7000 | -0.31141 | -3046.90000 | |
| 0 | -0.31937 | -583.19000 | -0.32013 | -117.16000 | -0.31008 | -3001.00000 | -0.30663 | -3115.10000 | -0.32607 |
| 0 | | | | | | | | | |

Table 18 : ε- Table for f.c.c lattice

| $\varepsilon_0^j$ | $\varepsilon_1^j$ | $\varepsilon_2^j$ | $\varepsilon_3^j$ | $\varepsilon_4^j$ | $\varepsilon_5^j$ | $\varepsilon_6^j$ | $\varepsilon_7^j$ | $\varepsilon_8^j$ | $\varepsilon_9^j$ |
|---|---|---|---|---|---|---|---|---|---|
| 0 | | | | | | | | | |
| 0 | -0.30231 | | | | | | | | |
| 0 | -0.26167 | 24.61000 | | | | | | | |
| 0 | -0.22952 | 31.10200 | -0.10764 | | | | | | |
| 0 | -0.21126 | 54.76000 | -0.18725 | 18.54100 | | | | | |
| 0 | -0.20993 | 752.88000 | -0.20983 | 10.46300 | -0.31105 | | | | |
| 0 | -0.22567 | -63.54100 | -0.21116 | 0.40527 | -0.30925 | 567.18000 | | | |
| 0 | -0.25573 | -33.26900 | -0.19264 | -9.54600 | -0.31164 | -417.60000 | -0.31027 | | |
| 0 | -0.29494 | -25.50100 | -0.12700 | -18.03400 | -0.31045 | 825.97000 | -0.31084 | -2163.80000 | |
| 0 | -0.33664 | 23.98000 | 0.36268 | -23.45900 | -0.31134 | -1132.90000 | -0.31096 | -7664.50000 | -0.31102 |
| 0 | | | | | | | | | |

# Table 19: ε- Table for s.c lattice

| $\varepsilon_0^j$ | $\varepsilon_1^j$ | $\varepsilon_2^j$ | $\varepsilon_3^j$ | $\varepsilon_4^j$ | $\varepsilon_5^j$ | $\varepsilon_6^j$ | $\varepsilon_7^j$ | $\varepsilon_8^j$ | $\varepsilon_9^j$ |
|---|---|---|---|---|---|---|---|---|---|
| 0 | | | | | | | | | |
| | -0.30325 | | | | | | | | |
| 0 | | -995.80000 | | | | | | | |
| | -0.30426 | | -0.30597 | | | | | | |
| 0 | | -1580.40000 | | -215.04000 | | | | | |
| | -0.30489 | | -0.30524 | | -0.30538 | | | | |
| 0 | | -4471.50000 | | -7301.70000 | | -2199.10000 | | | |
| | -0.30511 | | -0.30559 | | -0.30518 | | -0.32078 | | |
| 0 | | -6574.10000 | | -4852.30000 | | -2263.30000 | | -2314.90000 | |
| | -0.30527 | | -0.30501 | | -0.30479 | | -0.34016 | | -0.32579 |
| 0 | | -2688.70000 | | -171.07000 | | -2291.50000 | | -2245.30000 | |
| | -0.30564 | | -0.30461 | | -0.30527 | | -0.31855 | | |
| 0 | | -1714.80000 | | -1697.60000 | | -2366.80000 | | | |
| | -0.30622 | | -0.24620 | | -0.30676 | | | | |
| 0 | | -1698.10000 | | -1714.20000 | | | | | |
| | -0.30681 | | -0.30849 | | | | | | |
| 0 | | -2292.00000 | | | | | | | |
| | -0.30725 | | | | | | | | |
| 0 | | | | | | | | | |

# Table 20: ε- Table for d. lattice

| $\varepsilon_0^j$ | $\varepsilon_1^j$ | $\varepsilon_2^j$ | $\varepsilon_3^j$ | $\varepsilon_4^j$ | $\varepsilon_5^j$ |
|---|---|---|---|---|---|
| 0 | | | | | |
| | -0.087212 | | | | |
| 0 | | -18.62500 | | | |
| | -0.140904 | | -0.942676 | | |
| 0 | | -19.87230 | | -18.29190 | |
| | -0.191225 | | -0.309907 | | -0.326636 |
| 0 | | -28.29820 | | -78.066700 | |
| | -0.226563 | | -0.33000 | | |
| 0 | | -37.96600 | | | |
| | -0.252903 | | | | |
| 0 | | | | | |

**(B) Analysis of high temperature susceptibility series**

**(i) Estimation of the critical temperatures using the variable $\omega$ as $\tanh K$**

Domb [12] has provided the series expansions for the dimensionless suscptibility $\chi$ various lattices. The methodology of obtaining the critical temperatures using EA is entirely identical with the earlier one adopted for magnetization. Here too, we form the series for $\frac{d}{d\omega}\ln(\chi)$ where $\chi$ denotes the susceptibility and $\omega = \tanh K$.

$$\frac{d}{d\omega}\ln(\chi) = \sum_{n=0}^{N} g_n \omega^n \qquad (6)$$

Table 21 provides the series coefficients $\{g_n\}$ for b.c.c, f.c.c, s.c and d. lattices.

# Table 21: Coefficients $g_n$ for three-dimensional lattices (cf. Eqn 6)

| | $g_n$ | | | |
|---|---|---|---|---|
| **n** | **b.c.c** | **f.c.c** | **s.c** | **d.** |
| 0 | 8 | 12 | 6 | 4 |
| 1 | 48 | 120 | 24 | 8 |
| 2 | 344 | 1188 | 126 | 28 |
| 3 | 2016 | 11664 | 528 | 80 |
| 4 | 13928 | 114492 | 2646 | 244 |
| 5 | 83376 | 1124856 | 11160 | 584 |
| 6 | 567512 | 11057268 | 54942 | 1852 |
| 7 | 3443136 | 108689568 | 236448 | 5024 |
| 8 | 23173256 | 1068318972 | 1147590 | 14932 |
| 9 | 141900528 | 10500476760 | 4997304 | 38888 |
| 10 | 947483864 | 103210418532 | 24016878 | 118012 |
| 11 | 5840905824 | 1014487790928 | 105409872 | 318608 |
| 12 | 38777405480 | 9971896250796 | 503177382 | 941620 |
| 13 | 240242653104 | 98020011495288 | 2221205976 | 2526728 |
| 14 | 1588076108024 | 963511072360548 | 10550588046 | 7490428 |
| 15 | 9876516866432 | | 46779283008 | 20348480 |
| 16 | 65066890964808 | | 221344824198 | 59749156 |
| 17 | 405890414289648 | | 984810774456 | 162366632 |
| 18 | 2666799994767672 | | 4645471739406 | 476151100 |
| 19 | 16676604689911136 | | 20726801667408 | 1301565200 |
| 20 | 109326847463932232 | | 97524764573334 | 3799448884 |
| 21 | 685059337227459440 | | 436137683145144 | |
| 22 | 4482750380953107480 | | 2047836611612718 | |

In the ε-table, the ratios $\left\{\frac{g_n}{g_{n+1}}\right\}$ were computed using the $g_n$ coefficients of equation (6). Tables 22, 23, 24, and 25 provide these values for the lattices b.c.c ($g_1$ to $g_8$), f.c.c ($g_6$ to $g_{13}$), s.c ($g_{15}$ to $g_{22}$), and d. ($g_{15}$ to $g_{20}$), respectively.

Table 22: ε- Table for b.c.c lattice

| $\varepsilon_0^j$ | $\varepsilon_1^j$ | $\varepsilon_2^j$ | $\varepsilon_3^j$ | $\varepsilon_4^j$ | $\varepsilon_5^j$ | $\varepsilon_6^j$ | $\varepsilon_7^j$ |
|---|---|---|---|---|---|---|---|
| 0 | | | | | | | |
| | 0.139534 | | | | | | |
| 0 | | 32.154302 | | | | | |
| | 0.170634 | | 0.156506 | | | | |
| 0 | | -38.624174 | | 4495.073219 | | | |
| | 0.144744 | | 0.156726 | | 0.156606 | | |
| 0 | | 44.830845 | | -3815.027655 | | -7950.234930 | |
| | 0.167050 | | 0.156467 | | 0.156364 | | 0.15608 |
| 0 | | -49.663492 | | 13515.809507 | | -11577.432681 | |
| | 0.146914 | | 0.156393 | | 0.156880 | | |
| 0 | | 55.837207 | | -11462.722353 | | | |
| | 0.164824 | | 0.156306 | | | | |
| 0 | | -61.569355 | | | | | |
| | 0.148582 | | | | | | |
| 0 | | | | | | | |

Table 23 : ε- Table for f.c.c lattice

| $\varepsilon_0^j$ | $\varepsilon_1^j$ | $\varepsilon_2^j$ | $\varepsilon_3^j$ | $\varepsilon_4^j$ | $\varepsilon_5^j$ | $\varepsilon_6^j$ | $\varepsilon_7^j$ |
|---|---|---|---|---|---|---|---|
| 0 | | | | | | | |
| | 0.101732 | | | | | | |
| 0 | | 158408.840371 | | | | | |
| | 0.101738 | | 0.101740 | | | | |
| 0 | | 853517.157811 | | -225631.569547 | | | |
| | 0.101740 | | 0.101739 | | 0.101735 | | |
| 0 | | -662428.545481 | | -456331.463498 | | -716787.802627 | |
| | 0.101738 | | 0.101744 | | 0.101731 | | 0.10172 |
| 0 | | -487052.248525 | | -533100.211695 | | -913839.652489 | |
| | 0.101736 | | 0.101722 | | 0.101728 | | |
| 0 | | -558678.857225 | | -368189.394746 | | | |
| | 0.101734 | | 0.101727 | | | | |
| 0 | | -703151.603264 | | | | | |
| | 0.101733 | | | | | | |
| 0 | | | | | | | |

**Table 24: ε- Table for s.c lattice**

| $\varepsilon_0^j$ | $\varepsilon_1^j$ | $\varepsilon_2^j$ | $\varepsilon_3^j$ | $\varepsilon_4^j$ | $\varepsilon_5^j$ | $\varepsilon_6^j$ | $\varepsilon_7^j$ |
|---|---|---|---|---|---|---|---|
| 0 | 0.211341 | | | | | | |
| 0 | 0.224758 | 74.529392 | 0.218217 | | | | |
| 0 | 0.211993 | -78.339034 | 0.218214 | -412058.126274 | 0.218217 | | |
| 0 | 0.224128 | 82.406104 | 0.218197 | -59514.236843 | 0.218195 | -105398.400760 | 0.21813 |
| 0 | 0.212528 | -86.205927 | 0.218196 | -518647.537106 | 0.218198 | -121426.202739 | |
| 0 | 0.223610 | 90.240982 | 0.218183 | -77707.806962 | | | |
| 0 | 0.212974 | -94.027421 | | | | | |
| 0 | | | | | | | |

**Table 25: ε- Table for d. lattice**

| $\varepsilon_0^j$ | $\varepsilon_1^j$ | $\varepsilon_2^j$ | $\varepsilon_3^j$ | $\varepsilon_4^j$ | $\varepsilon_5^j$ |
|---|---|---|---|---|---|
| 0 | 0.340565 | | | | |
| 0 | 0.367989 | 36.464428 | 0.354386 | | |
| 0 | 0.340998 | -37.049376 | 0.353931 | -2234.887519 | 0.35378 |
| 0 | 0.365829 | 40.271439 | 0.353818 | -8855.727375 | |
| 0 | 0.342566 | -42.987128 | | | |
| 0 | | | | | |

**(ii) Estimation of the critical temperatures using the variable $\omega_1 = 1 - e^{-4K}$**

It is of interest to enquire whether $\omega(= \tanh K)$ occurring in the conventional susceptibility expansions can be replaced by $\omega_1(= 1 - e^{-4K})$. The relation governing the two variables is given by $\omega = \frac{(1-\omega_1)^{-\frac{1}{2}}-1}{(1-\omega_1)^{-\frac{1}{2}}+1}$

Hence

$$\frac{d\{ln(\chi)\}}{d\,\omega_1} = \sum_{n=0}^{N} h_n\, \omega_1^n \qquad (7)$$

The coefficients $\{h_n\}$ are listed in Table 26, for the three-dimensional lattices.

**Table 26: Coefficients $h_n$ for three-dimensional lattices (cf. Eqn 7)**

| | $h_n$ | | | |
|---|---|---|---|---|
| n | b.c.c | f.c.c | s.c | d. |
| 0 | 2 | 3 | 1.5 | 1 |
| 1 | 5 | 10.5 | 3 | 1.5 |
| 2 | 11.75 | 32.625 | 5.625 | 2.125 |
| 3 | 25.625 | 98.4375 | 9.9375 | 2.9375 |
| 4 | 55.671875 | 294.4453125 | 17.3203125 | 4.0078 |
| 5 | 119.46875 | 878.625 | 29.8125 | 5.3906 |
| 6 | 256.3789063 | 2620.555664 | 51.046875 | 7.166 |
| 7 | 548.9462891 | 7815.856445 | 87.1069336 | 9.4468 |
| 8 | 1175.576172 | 23311.83203 | 148.400848 | 12.385 |
| 9 | 2516.383301 | 69531.85938 | 252.595642 | 16.181 |
| 10 | 5386.781738 | 207393.0313 | 429.761505 | 21.089 |
| 11 | 11530.24023 | 618593.75 | 731.023376 | 27.441 |
| 12 | 24680.57422 | 1845097 | 1243.33313 | 35.662 |
| 13 | 52827.49219 | 5503564 | 2114.56421 | 46.305 |
| 14 | 113075.3125 | 16415688 | 3596.17773 | 60.083 |
| 15 | 242037.7813 | | 6115.95215 | 77.923 |
| 16 | 518082.625 | | 10401.2598 | 101.02 |
| 17 | 1108981.25 | | 17689.7461 | 130.93 |
| 18 | 2374038 | | 30085.4766 | 169.67 |
| 19 | 5081659 | | 51165.9609 | 219.83 |
| 20 | 10865928 | | 87009.625 | 284.77 |
| 21 | 23254136 | | 148035.125 | 368.94 |
| 22 | 49739312 | | 251850 | 477.99 |
| 23 | 107113536 | | 428163.762 | |
| 24 | 229309560 | | 727730.227 | |
| 25 | | | 1240957.64 | |
| 26 | | | 2120726.44 | |

The ratios $\left\{\frac{h_n}{h_{n+1}}\right\}$, were computed and are shown for the lattices b.c.c ($h_{17}$ to $h_{24}$), f.c.c ($h_7$ to $h_{14}$), s.c ($h_{17}$ to $h_{24}$), and d. ($h_{14}$ to $h_{21}$) in Tables 27, 28, 29, and 30, respectively. The prescription of EA given by equation (1) is employed, as before to obtain the respective limits.

**Table 27: ε- Table for b.c.c lattice**

| $\varepsilon_0^j$ | $\varepsilon_1^j$ | $\varepsilon_2^j$ | $\varepsilon_3^j$ | $\varepsilon_4^j$ | $\varepsilon_5^j$ | $\varepsilon_6^j$ | $\varepsilon_7^j$ |
|---|---|---|---|---|---|---|---|
| 0 | | | | | | | |
| | 0.46712 | | | | | | |
| 0 | | 20383.89000 | | | | | |
| | 0.46717 | | 0.46712 | | | | |
| 0 | | 2035.09700 | | 5109.92400 | | | |
| | 0.46766 | | 0.46744 | | 0.46742 | | |
| 0 | | -2496.86500 | | -42013.35000 | | 3865.21400 | |
| | 0.46726 | | 0.46742 | | 0.46744 | | 0.46715 |
| 0 | | 3973.93700 | | -3380.25800 | | 482.40080 | |
| | 0.46752 | | 0.46728 | | 0.46770 | | |
| 0 | | -316.49560 | | -1003.55500 | | | |
| | 0.46436 | | 0.46583 | | | | |
| 0 | | 363.29260 | | | | | |
| | 0.46711 | | | | | | |
| 0 | | | | | | | |

**Table 28: ε- Table for f.c.c lattice**

| $\varepsilon_0^j$ | $\varepsilon_1^j$ | $\varepsilon_2^j$ | $\varepsilon_3^j$ | $\varepsilon_4^j$ | $\varepsilon_5^j$ | $\varepsilon_6^j$ | $\varepsilon_7^j$ |
|---|---|---|---|---|---|---|---|
| 0 | | | | | | | |
| | 0.335274 | | | | | | |
| 0 | | -170219.100000 | | | | | |
| | 0.335268 | | 0.335264 | | | | |
| 0 | | -450395.100000 | | 14462690.000000 | | | |
| | 0.335266 | | 0.335264 | | 0.335264 | | |
| 0 | | -1231355.000000 | | -721990.100000 | | 532381.200000 | |
| | 0.335265 | | 0.335266 | | 0.335265 | | 0.33526 |
| 0 | | -561580.500000 | | -1534173.000000 | | -525016.800000 | |
| | 0.335263 | | 0.335265 | | 0.335266 | | |
| 0 | | -116256.200000 | | -262542.50000 | | | |
| | 0.335254 | | 0.335258 | | | | |
| 0 | | 132888.800000 | | | | | |
| | 0.335262 | | | | | | |
| 0 | | | | | | | |

**Table 29: ε- Table for s.c lattice**

| $\varepsilon_0^j$ | $\varepsilon_1^j$ | $\varepsilon_2^j$ | $\varepsilon_3^j$ | $\varepsilon_4^j$ | $\varepsilon_5^j$ | $\varepsilon_6^j$ | $\varepsilon_7^j$ |
|---|---|---|---|---|---|---|---|
| 0 | | | | | | | |
| | 0.58798 | | | | | | |
| 0 | | 66774.99000 | | | | | |
| | 0.58799 | | 0.58797 | | | | |
| 0 | | 19398.43000 | | 53910.57000 | | | |
| | 0.58804 | | 0.58800 | | 0.58798 | | |
| 0 | | -3495.80000 | | -8097.49800 | | -23015.5500 | |
| | 0.58776 | | 0.58778 | | 0.58792 | | 0.58794 |
| 0 | | 36402.96000 | | -643.05570 | | 16017.54000 | |
| | 0.58779 | | 0.58776 | | 0.58798 | | |
| 0 | | 2388.51300 | | 3878.43100 | | | |
| | 0.58820 | | 0.58843 | | | | |
| 0 | | 6870.62900 | | | | | |
| | 0.58835 | | | | | | |
| 0 | | | | | | | |

**Table 30: ε- Table for d. lattice**

| $\varepsilon_0^j$ | $\varepsilon_1^j$ | $\varepsilon_2^j$ | $\varepsilon_3^j$ | $\varepsilon_4^j$ | $\varepsilon_5^j$ | $\varepsilon_6^j$ | $\varepsilon_7^j$ |
|---|---|---|---|---|---|---|---|
| 0 | | | | | | | |
| | 0.771064 | | | | | | |
| 0 | | 3472.82500 | | | | | |
| | 0.77135 | | 0.77207 | | | | |
| 0 | | 4862.25600 | | -17687.77000 | | | |
| | 0.77155 | | 0.77202 | | 0.77206 | | |
| 0 | | 6991.23500 | | 13071.04000 | | -191.58350 | |
| | 0.77170 | | 0.77219 | | 0.77198 | | 0.77207 |
| 0 | | 9027.28900 | | 8248.54700 | | 10923.75000 | |
| | 0.77181 | | 0.77090 | | 0.77235 | | |
| 0 | | 7920.78700 | | 8937.93400 | | | |
| | 0.77193 | | 0.77189 | | | | |
| 0 | | -13420.43000 | | | | | |
| | 0.77186 | | | | | | |
| 0 | | | | | | | |

**(iii) Estimation of the critical exponent γ using the variable $\omega$**

The methodology for the estimation of γ here is identical with that outlined earlier. Thus, we form the new series

$$ln(\chi) = \sum_{n=0}^{N} l_n \Omega^n \qquad (8)$$

where $\Omega = \frac{\omega}{\omega_c}$. Furthermore, employing $\omega_c$ as 0.156085, 0.101707, 0.218090 and 0.353729 [6,7] respectively for b.c.c, f.c.c, s.c and d. lattices. The coefficients $\{l_n\}$ are shown in Table 31 for the four lattices.

**Table 31 : Coefficients $l_n$ for three-dimensional lattices (cf. Eqn 8)**

| | $l_n$ | | | |
|---|---|---|---|---|
| n | b.c.c | f.c.c | s.c | d. |
| 0 | 0 | 0 | 0 | 0 |
| 1 | 1.24868 | 1.2204853 | 1.30854 | 1.414917111 |
| 2 | 0.58470065 | 0.62066017 | 0.570758977 | 0.500497607 |
| 3 | 0.43603434 | 0.41662864 | 0.435668889 | 0.413094867 |
| 4 | 0.2991405 | 0.31202743 | 0.298618659 | 0.313122319 |
| 5 | 0.25806232 | 0.24920784 | 0.261095026 | 0.270255697 |
| 6 | 0.20093557 | 0.20751694 | 0.200137034 | 0.190672315 |
| 7 | 0.18298066 | 0.17783162 | 0.184186032 | 0.183332725 |
| 8 | 0.15161897 | 0.15556374 | 0.15126264 | 0.153931571 |
| 9 | 0.14157777 | 0.13823571 | 0.142320027 | 0.143851547 |
| 10 | 0.12178553 | 0.12437186 | 0.121644729 | 0.119268479 |
| 11 | 0.1153857 | 0.11303031 | 0.115908996 | 0.116389704 |
| 12 | 0.1017731 | 0.10358129 | 0.101701901 | 0.101889051 |
| 13 | 0.097348797 | 0.095587579 | 0.097733295 | 0.098323067 |
| 14 | 0.087413486 | 0.088737071 | 0.087369798 | 0.086661172 |
| 15 | 0.084177802 | | 0.084473734 | 0.084816527 |
| 16 | 0.076605954 | | 0.076578282 | 0.076409632 |
| 17 | 0.074139595 | | 0.074375216 | 0.074694733 |
| 18 | 0.06817627 | | 0.068159097 | 0.067811353 |
| 19 | 0.066236755 | | 0.066431046 | 0.06664082 |
| 20 | 0.061418755 | | 0.061401367 | 0.06121475 |
| 21 | 0.059853848 | | 0.060013149 | 0.06019944 |
| 22 | 0.055664063 | | 0.055664063 | |
| 23 | 0.054443359 | | 0.055664063 | |
| 24 | 0.051269531 | | 0.051250191 | |

For a typical ε-table, the sequence $\{n\, l_n\}$ was calculated using the $l_n$ coefficients for b.c.c ($l_{18}$ to $l_{24}$), f.c.c ($l_6$ to $l_{12}$), s.c ($l_{10}$ to $l_{16}$), and d. ($l_1$ to $l_7$) lattices and are shown in Tables 32, 33, 34, and 35, respectively.

**Table 32: ε- Table for b.c.c lattice**

| $\varepsilon_0^j$ | $\varepsilon_1^j$ | $\varepsilon_2^j$ | $\varepsilon_3^j$ | $\varepsilon_4^j$ | $\varepsilon_5^j$ | $\varepsilon_6^j$ | $\varepsilon_7^j$ |
|---|---|---|---|---|---|---|---|
| 0 | | | | | | | |
| | 1.22718 | | | | | | |
| 0 | | 31.93466 | | | | | |
| | 1.25849 | | 1.24314 | | | | |
| 0 | | -33.19695 | | -9087.16273 | | | |
| | 1.22837 | | 1.24303 | | 1.24362 | | |
| 0 | | 35.01927 | | -7390.72207 | | -8267.75435 | |
| | 1.25693 | | 1.24289 | | 1.24248 | | 1.24044 |
| 0 | | -36.25095 | | -9794.14815 | | -8757.92738 | |
| | 1.22934 | | 1.24279 | | 1.24344 | | |
| 0 | | 38.08802 | | -8259.11965 | | | |
| | 1.25560 | | 1.24267 | | | | |
| 0 | | -39.28978 | | | | | |
| | 1.23014 | | | | | | |
| 0 | | | | | | | |

**Table 33: ε- Table for f.c.c lattice**

| $\varepsilon_0^j$ | $\varepsilon_1^j$ | $\varepsilon_2^j$ | $\varepsilon_3^j$ | $\varepsilon_4^j$ | $\varepsilon_5^j$ | $\varepsilon_6^j$ | $\varepsilon_7^j$ |
|---|---|---|---|---|---|---|---|
| 0 | | | | | | | |
| | 1.24510 | | | | | | |
| 0 | | -3567.59086 | | | | | |
| | 1.24482 | | 1.24762 | | | | |
| 0 | | -3211.13261 | | -3857.17178 | | | |
| | 1.24450 | | 1.24607 | | 1.24679 | | |
| 0 | | -2753.74354 | | -2463.37769 | | -2633.07594 | |
| | 1.24412 | | 1.25513 | | 1.24090 | | 1.23752 |
| 0 | | -2482.98472 | | -2533.62154 | | -2629.17417 | |
| | 1.24371 | | 1.23539 | | 1.23837 | | |
| 0 | | -2603.06849 | | -2198.52846 | | | |
| | 1.24333 | | 1.23789 | | | | |
| 0 | | -2785.83658 | | | | | |
| | 1.24297 | | | | | | |
| 0 | | | | | | | |

**Table 34: ε- Table for s.c lattice**

| $\varepsilon_0^j$ | $\varepsilon_1^j$ | $\varepsilon_2^j$ | $\varepsilon_3^j$ | $\varepsilon_4^j$ | $\varepsilon_5^j$ | $\varepsilon_6^j$ | $\varepsilon_7^j$ |
|---|---|---|---|---|---|---|---|
| 0 | | | | | | | |
| | 1.21644 | | | | | | |
| 0 | | 17.07893 | | | | | |
| | 1.27499 | | 1.24675 | | | | |
| 0 | | -18.32302 | | -4891.37933 | | | |
| | 1.22042 | | 1.24654 | | 1.24701 | | |
| 0 | | 19.95608 | | -2751.30929 | | -3477.34797 | |
| | 1.27053 | | 1.24618 | | 1.24563 | | 1.24152 |
| 0 | | -21.11679 | | -4572.08434 | | -3720.49857 | |
| | 1.22317 | | 1.24596 | | 1.24681 | | |
| 0 | | 22.76408 | | -3388.36754 | | | |
| | 1.26710 | | 1.24567 | | | | |
| 0 | | -23.89286 | | | | | |
| | 1.22525 | | | | | | |
| 0 | | | | | | | |

**Table 35: ε- Table for d. lattice**

| $\varepsilon_0^j$ | $\varepsilon_1^j$ | $\varepsilon_2^j$ | $\varepsilon_3^j$ | $\varepsilon_4^j$ | $\varepsilon_5^j$ | $\varepsilon_6^j$ | $\varepsilon_7^j$ |
|---|---|---|---|---|---|---|---|
| 0 | | | | | | | |
| | 1.41491 | | | | | | |
| 0 | | -2.41591 | | | | | |
| | 1.00099 | | 1.15222 | | | | |
| 0 | | 4.19657 | | 14.09366 | | | |
| | 1.23928 | | 1.25326 | | 0 | | |
| 0 | | 75.73075 | | 13.29574 | | 14.06934 | |
| | 1.25248 | | 1.23724 | | 1.29266 | | 1.24238 |
| 0 | | 10.12256 | | 31.33973 | | -5.81870 | |
| | 1.35127 | | 1.28437 | | 1.26575 | | |
| 0 | | -4.82521 | | -22.35647 | | | |
| | 1.14403 | | 1.22733 | | | | |
| 0 | | 7.17899 | | | | | |
| | 1.28332 | | | | | | |
| 0 | | | | | | | |

**(iv) Estimation of the exponent γ using the variable $\omega_1$**

Employing $(\omega_1)_c = 1 - e^{-4K_c}$ = 0.46713, 0.33519, 0.58794 and 0.77210 for b.c.c, f.c.c, s.c and d. lattices [6,7], the new series for $\ln(\chi)$ may be written as

$$\ln(\chi) = \sum_{n=0}^{N} m_n \Omega_1^n \qquad (9)$$

where $\Omega_1 = \frac{\omega_1}{(\omega_1)_c}$. The values of $K_c$ are from [6,7]. The coefficients $\{m_n\}$ are shown in Table 36 for the four lattices. the sequence $\{nm_n\}$ is utilized to calculate the second column in the ε- table.

**Table 36: Coefficients $m_n$ for three-dimensional lattices (cf. Eqn 9)**

| | $m_n$ | | | |
|---|---|---|---|---|
| **n** | **b.c.c** | **f.c.c** | **s.c** | **d.** |
| 0 | 0 | 0 | 0 | 0 |
| 1 | 0.93427 | 1.0056 | 0.88192 | 0.77211 |
| 2 | 0.545529784 | 0.58988432 | 0.518518557 | 0.447113074 |
| 3 | 0.399213336 | 0.409529952 | 0.381074401 | 0.326033466 |
| 4 | 0.30509468 | 0.310632294 | 0.296868596 | 0.26099885 |
| 5 | 0.247649849 | 0.249187759 | 0.243412844 | 0.219949524 |
| 6 | 0.206883409 | 0.207677958 | 0.205173656 | 0.190350078 |
| 7 | 0.177834928 | 0.178030158 | 0.177095249 | 0.167464772 |
| 8 | 0.15554606 | 0.155668968 | 0.155519126 | 0.149148914 |
| 9 | 0.138338869 | 0.138318185 | 0.138453521 | 0.13419956 |
| 10 | 0.124457472 | 0.124495249 | 0.124672759 | 0.121833711 |
| 11 | 0.113250111 | 0.113177786 | 0.113387255 | 0.111463591 |
| 12 | 0.103678312 | 0.103721286 | 0.103961952 | 0.102642996 |
| 13 | 0.09582774 | 0.09565768 | 0.09590882 | 0.095081237 |
| 14 | 0.088868526 | 0.088927696 | 0.089043798 | 0.088508299 |
| 15 | 0.082960773 | 0.082844092 | 0.083180634 | 0.082759066 |
| 16 | 0.077787431 | | 0.077972319 | 0.077698987 |
| 17 | 0.073125951 | | 0.073278674 | 0.073192412 |
| 18 | 0.069179042 | | 0.069309269 | 0.069178852 |
| 19 | 0.065404356 | | 0.065534679 | 0.065575805 |
| 20 | 0.062208385 | | 0.062341955 | 0.062316192 |
| 21 | 0.059286484 | | 0.059340472 | 0.059412555 |
| 22 | 0.056481638 | | 0.056652887 | 0.056631037 |
| 23 | 0.054053225 | | 0.054166369 | 0.054257946 |
| 24 | 0.051805838 | | 0.051848453 | 0.05190475 |
| 25 | 0.049703384 | | 0.049877205 | 0.049775809 |
| 26 | 0.046525608 | | 0.047760579 | 0.047695424 |
| 27 | | | 0.04552415 | |
| 28 | | | 0.041864407 | |

The ε- table coefficients $\{nm_n\}$ for the lattices b.c.c ($m_{10}$ to $m_{16}$), f.c.c ($m_7$ to $m_{13}$), s.c ($m_{17}$ to $m_{23}$), and d. ($m_{13}$ to $m_{19}$), are shown in Tables 37, 38, 39, and 40, respectively.

**Table 37: ε- Table for b.c.c lattice**

| $\varepsilon_0^j$ | $\varepsilon_1^j$ | $\varepsilon_2^j$ | $\varepsilon_3^j$ | $\varepsilon_4^j$ | $\varepsilon_5^j$ | $\varepsilon_6^j$ | $\varepsilon_7^j$ |
|---|---|---|---|---|---|---|---|
| 0 | | | | | | | |
| | 1.24457 | | | | | | |
| 0 | | 849.98500 | | | | | |
| | 1.24575 | | 1.24507 | | | | |
| 0 | | -620.55100 | | -8726.46500 | | | |
| | 1.24414 | | 1.24494 | | 1.24495 | | |
| 0 | | 616.95120 | | 137573.80000 | | -11889.13000 | |
| | 1.24576 | | 1.24495 | | 1.24494 | | 1.24508 |
| 0 | | -624.5067 | | -2355.00900 | | -4629.34600 | |
| | 1.24415 | | 1.24437 | | 1.24450 | | |
| 0 | | 3964.60500 | | 5276.88800 | | | |
| | 1.24441 | | 1.24513 | | | | |
| 0 | | 5338.81200 | | | | | |
| | 1.24459 | | | | | | |
| 0 | | | | | | | |

## Table 38: ε- Table for f.c.c lattice

| $\varepsilon_0^j$ | $\varepsilon_1^j$ | $\varepsilon_2^j$ | $\varepsilon_3^j$ | $\varepsilon_4^j$ | $\varepsilon_5^j$ | $\varepsilon_6^j$ | $\varepsilon_7^j$ |
|---|---|---|---|---|---|---|---|
| 0 |  |  |  |  |  |  |  |
|   | 1.246211 |  |  |  |  |  |  |
| 0 |  | -1163.66020 |  |  |  |  |  |
|   | 1.245351 |  | 1.244222 |  |  |  |  |
| 0 |  | -2048.83630 |  | -653.98234 |  |  |  |
|   | 1.244863 |  | 1.244938 |  | 1.244953 |  |  |
| 0 |  | 11234.19400 |  | 70760.85000 |  | 5272.41850 |  |
|   | 1.244952 |  | 1.244955 |  | 1.244939 |  | 1.24503 |
| 0 |  | 337018.78000 |  | 8053.75780 |  | 5280.60700 |  |
|   | 1.244955 |  | 1.244952 |  | 1.244579 |  |  |
| 0 |  | -3330.94000 |  | 5376.43980 |  |  |  |
|   | 1.244655 |  | 1.245067 |  |  |  |  |
| 0 |  | -904.47632 |  |  |  |  |  |
|   | 1.243549 |  |  |  |  |  |  |
| 0 |  |  |  |  |  |  |  |

## Table 39: ε- Table for s.c lattice

| $\varepsilon_0^j$ | $\varepsilon_1^j$ | $\varepsilon_2^j$ | $\varepsilon_3^j$ | $\varepsilon_4^j$ | $\varepsilon_5^j$ | $\varepsilon_6^j$ | $\varepsilon_7^j$ |
|---|---|---|---|---|---|---|---|
| 0 |  |  |  |  |  |  |  |
|   | 1.24573 |  |  |  |  |  |  |
| 0 |  | 546.63160 |  |  |  |  |  |
|   | 1.24756 |  | 1.24652 |  |  |  |  |
| 0 |  | -415.29300 |  | -3056.64600 |  |  |  |
|   | 1.24515 |  | 1.24614 |  | 1.24631 |  |  |
| 0 |  | 595.04220 |  | 3083.78200 |  | 10386.96000 |  |
|   | 1.24683 |  | 1.24655 |  | 1.24644 |  | 1.24636 |
| 0 |  | -2865.45100 |  | -6717.63300 |  | -928.39370 |  |
|   | 1.24649 |  | 1.24629 |  | 1.24662 |  |  |
| 0 |  | -7875.70300 |  | -3690.03200 |  |  |  |
|   | 1.24636 |  | 1.24653 |  |  |  |  |
| 0 |  | -1862.16900 |  |  |  |  |  |
|   | 1.24582 |  |  |  |  |  |  |
| 0 |  |  |  |  |  |  |  |

## Table 40: ε- Table for d. lattice

| $\varepsilon_0^j$ | $\varepsilon_1^j$ | $\varepsilon_2^j$ | $\varepsilon_3^j$ | $\varepsilon_4^j$ | $\varepsilon_5^j$ | $\varepsilon_6^j$ | $\varepsilon_7^j$ |
|---|---|---|---|---|---|---|---|
| 0 |  |  |  |  |  |  |  |
|   | 1.23605 |  |  |  |  |  |  |
| 0 |  | 326.79120 |  |  |  |  |  |
|   | 1.23911 |  | 1.24790 |  |  |  |  |
| 0 |  | 440.55790 |  | 911.17360 |  |  |  |
|   | 1.24138 |  | 1.25003 |  | 1.24836 |  |  |
| 0 |  | 556.23220 |  | 312.08460 |  | -529.31090 |  |
|   | 1.24318 |  | 1.24593 |  | 1.24717 |  | 1.24762 |
| 0 |  | 919.70270 |  | 1119.71500 |  | 1709.15500 |  |
|   | 1.24427 |  | 1.25093 |  | 1.24887 |  |  |
| 0 |  | 1069.77100 |  | 635.44860 |  |  |  |
|   | 1.24520 |  | 1.24863 |  |  |  |  |
| 0 |  | 1361.62100 |  |  |  |  |  |
|   | 1.24594 |  |  |  |  |  |  |
| 0 |  |  |  |  |  |  |  |

**Results and Discussion**

The foregoing analysis has demonstrated the efficacy of EA to deduce the critical temperatures and exponents for series expansions pertaining to magnetization as well as susceptibility. Since the series coefficients arising in three-dimensional Ising models do not follow a systematic pattern, the numerical analysis is rendered difficult. It is well-known that EA is closely related to the Pade' Approximants [30]. However, the former is easier to implement and has more flexibility in restricting or enlarging the number of coefficients of a given power series. The applicability of EA has been demonstrated for deducing the critical properties of a few Ising models [25] as well as for calculating the close packing densities pertaining to virial equations of state for hard spheres and hard disks [26].

We have estimated the critical temperatures (Table 41) and exponents β and γ (Table 42) for the simple cubic, body-centered cubic, face-centered cubic and diamond lattices. Wakefield [31] has reported $K_c$ as 0.222363 for s.c lattices nearly seven decades ago, using series expansions. Depending upon the computational methodology and system size, other values have been reported [32] For simple cubic lattices, Rosengren [33] has proposed the following expression for $K_c$ using the generating function approach viz

$$\tanh K_c = (\sqrt{5} - 2) \cos\left(\frac{\pi}{8}\right)$$

Several other estimates for $K_c$ have also been reported *viz* 0.221655[34], 0.22166[35], 0.221655[36], 0.22165434([37], 0.221650[38,39] using diverse strategies. Using accurate Monte Carlo simulations along with finite-size scaling for f.c.c lattices, Yu [40] has deduced $K_c$ as 0.1020707.

The EA algorithm has led to satisfactory critical temperatures for all the lattices, although the low temperature variable $x$ in the magnetization series for s.c lattice, was modified as $x \ln 2$. A plausible interpretation may be as follows: In deriving the low temperature series expansions, a factor of $\ln 2$ has been omitted while considering the thermodynamic limit viz $N \to \infty$. This may probably be the reason for including $\ln 2$ in the parametric definition of $x$.

EA *per se* seems applicable in extracting the critical temperatures if suitable ranges of coefficients are chosen. It is not our endeavor here to extract more exact critical temperatures by analyzing diverse variants of EA [41] but rather to point out the effectiveness of the original EA [23,24] in this context. A perusal of the initial non-zero entries for b.c.c (Table 2) and (f.c.c (Table 3) indicates that the ratios of the series coefficients do not follow any pattern in the sign and / or magnitude and the final limits could not have been anticipated. For the s.c lattices, only upon re-formulating the series expansion, the trend in the ratios of the coefficients (Table 3) became regular and EA led to the anticipated critical temperature. For the d. lattices, the coefficients exhibited a regular pattern and hence the estimation of $x_c$ became easier.

It is well known that Pade' Approximants [10] provide satisfactory estimates of the $K_c$ from magnetization series while the ratio method [30] is valid for diamond lattices. We emphasize that from the successive ratios of the coefficients, the critical exponent $\alpha$ from the series for specific heat has been determined. [42]. The dichotomy between the closest singularity to the origin and physical singularity is a crucial factor in any analysis of series expansions.

**Table 41- Estimates of $K_c$ from series expansions for (A)susceptibility and (B)magnetization**

| Lattice | $K_c$ using the variable | | $K_c$ using the variable | | Simulation data |
|---|---|---|---|---|---|
| | $x$ | $x_1$ | $\omega$ | $\omega_1$ | |
| f.c.c | 0.10247±0.00662 | 0.10218±0.00166 | 0.10207±0.000007 | 0.10209±0.000005 | 0.102065[6] |
| s.c | 0.22158±0.00018 | 0.22085±0.00118 | 0.22168±0.000018 | 0.22165±0.000035 | 0.221659[7] |
| b.c.c | 0.15476±0.00703 | 0.15768±0.00225 | 0.15737±0.000018 | 0.15737±0.00001 | 0.157371[6] |
| d. | 0.36998±0.00074 | 0.36959±0.00217 | 0.36974±0.000206 | 0.36973±0.000045 | 0.369720[6] |

**Table 42-Estimates of $\beta$ and $\gamma$ using two different variables**

| Lattice | $\beta$ using the variable | | $\gamma$ using the variable | |
|---|---|---|---|---|
| | $x$ | $x_1$ | $\omega$ | $\omega_1$ |
| f.c.c | 0.32661 ± 0.00084 | 0.31021±0.00082 | 1.23829±0.00241 | 1.24517±0.00025 |
| s.c | 0.326606 ± 0.00135 | 0.32574± 0.00147 | 1.24347±0.00251 | 1.24632±0.00058 |
| b.c.c | 0.32919 ±0.00169 | 0.32614±0.00280 | 1.24083±0.00045 | 1.24445±0.00045 |
| d. | 0.326612±0.01042 | 0.32121 ±0.00386 | 1.24499±0.00195 | 1.24762±0.00091 |

We have attempted herein, a systematic study on the influence of various coefficients in dictating the magnitude of the critical temperatures. The mean values and standard deviations of $K_c$ are reported in Table 41. In view of vary large number of *permutations* possible for each series, we have confined ourselves to a few coefficients ranges alone. It is gratifying to note that the $K_c$ values obtained here are entirely consistent with the simulation data and the agreement is not fortuitous. It is customary to analyze the magnetization series in terms of $e^{-4K}$ while Guttmann and Thompson [28] have emphasized that a more natural variable is $1 - \tanh K$. We have estimated $K_c$ using both variables as shown in Table 41 and we infer that for f.c.c lattices, enhanced agreement with simulation data arises if $1 - \tanh K$ is chosen. For other lattices, both variables yield nearly identical estimates of $K_c$. A corollary of the above variable transformation is the similar exercise pertaining to susceptibility series expansions. Hence, we replace the usual high-temperature variable $\tanh K$ with $1 - e^{-4K}$ and analyze the magnitude of $K_c$ ; for s.c lattices, use of $(1 - e^{-4K})$ yields a better agreement with simulation data and in other cases, the preferred variable is $\tanh K$, as far as the susceptibility series is concerned.

      The calculation of $\beta$ on the other hand was straightforward for all the lattices considered herein, without the need to alter *the* well-known series expansions. The estimated values are consistent with reported values, despite minor differences among various lattices. On the basis of the values of Table 42, it appears that the analysis using $(1 - \tanh K)$ yields a more accurate value, only for the b.c.c lattice. For other lattices, $e^{-4K}$ as the variable compares more favorably with the simulation data. Interestingly, the magnetization series was rewritten by altering the original variable $x$ with a multiplicative factor $\ln 2$, to obtain the correct value of the critical temperatures for s.c lattices. On the other hand, the original magnetization series itself was adequate to calculate β. This indicates that $K_c$ is more sensitive to the series coefficients, in contrast to $\beta$ when viewed from the perspective offered by EA.

      The precise value of the critical exponent $\gamma$ for three-dimensional Ising models has been an issue of much interest [ 8,9,20-22,43-45]. Fisher and Perk [45] have provided a detailed critique on the relation among various exponents. As shown in Table 42, the mean value of γ for f.c.c, s.c, b.c.c and d. lattices are respectively

1.23829, 1.24347, 1.24083 and 1.24499, while employing $\tanh K$ as the variable. If $(1 - e^{-4K})$ is used, much larger exponents arise indicating that $\tanh K$ is the most appropriate variable for susceptibility series. The $\gamma$ values deduced herein are consistent with the Monte Carlo simulation estimate (1.23719) of Hasenbush [8,9], light scattering study of Sengers and Shanks involving fluids (1.238 ±0.012) [46]. Interestingly, irrespective of the number of coefficients considered in the series expansion, $\gamma$ is < 1.25, while using EA. In the numerical evaluation of the critical temperatures and exponents of three-dimensional models, the analysis of the former is more tedious [47-50], while the latter can be independently obtained using scaling hypothesis and related conjectures.

**Summary**

The applicability of $\varepsilon$-convergence algorithm to estimate the critical temperatures of simple cubic, face-centered cubic, body-centered cubic and diamond lattices was analyzed. The series expansions for low temperature magnetization and high temperature susceptibility were analyzed using two different variables, for dimensionless temperature. The critical exponents β and γ were also estimated using the algorithm. The mean values of the critical parameters are compared with the Monte Carlo simulation data.

**Acknowledgements**

This work was supported by the Mathematical Research Impact Centric Scheme (MATRICS) of SERB, Government of India. The computations were carried out using Python programming and the facilities of the P.G. Senapathy Centre for Computing Resources, Indian Institute of Technology-Madras are gratefully acknowledged.

**Appendix**

In this Appendix, we indicate the range of coefficients chosen for estimating the mean values of the critical parameters.

**(A) Low temperature magnetization series**

**(i) Estimation of $K_c$**

The coefficients chosen are given below

| Range of coefficients | | | | | | | |
|---|---|---|---|---|---|---|---|
| Series coefficients using $x$ | | | | Series coefficients using $x_1$ | | | |
| b.c.c | f.c.c | s.c | d. | b.c.c | f.c.c | s.c | d. |
| $a_8$ to $a_{19}$ | $a_{25}$ to $a_{31}$ | $a_9$ to $a_{14}$ | $a_1$ to $a_{12}$ | $b_{17}$ to $b_{26}$ | $b_{23}$ to $b_{32}$ | $b_{12}$ to $b_{23}$ | $b_{15}$ to $b_{28}$ |
| $a_{15}$ to $a_{24}$ | $a_{25}$ to $a_{33}$ | $a_9$ to $a_{21}$ | $a_1$ to $a_6$ | $b_{12}$ to $b_{17}$ | $b_{21}$ to $b_{30}$ | $b_{15}$ to $b_{26}$ | $b_0$ to $b_9$ |
| $a_{10}$ to $a_{25}$ | $a_{16}$ to $a_{24}$ | $a_{15}$ to $a_{27}$ | $a_4$ to $a_{13}$ | $b_{12}$ to $b_{19}$ | $b_{19}$ to $b_{40}$ | $b_{11}$ to $b_{22}$ | $b_4$ to $b_{13}$ |
| - | $a_{25}$ to $a_{35}$ | $a_{19}$ to $a_{29}$ | $a_5$ to $a_{12}$ | $b_{15}$ to $b_{26}$ | $b_{18}$ to $b_{35}$ | - | $b_6$ to $b_{15}$ |
| - | $a_{29}$ to $a_{37}$ | $a_{18}$ to $a_{28}$ | $a_6$ to $a_{13}$ | - | - | - | $b_{15}$ to $b_{22}$ |
| - | - | - | $a_7$ to $a_{12}$ | - | - | - | $b_{15}$ to $b_{30}$ |

**(ii) Estimation of β**

The coefficients chosen are given below

| Range of coefficients | | | | | | | |
|---|---|---|---|---|---|---|---|
| Series coefficients using $x$ | | | | Series coefficients using $x_1$ | | | |
| b.c.c | f.c.c | s.c | d. | b.c.c | f.c.c | s.c | d. |
| $c_{20}$ to $c_{25}$ | $c_{15}$ to $c_{23}$ | $c_{11}$ to $c_{17}$ | $c_8$ to $c_{12}$ | $d_8$ to $d_{18}$ | $d_{71}$ to $d_{79}$ | $d_{48}$ to $d_{54}$ | $d_4$ to $d_8$ |
| $c_7$ to $c_{11}$ | $c_{17}$ to $c_{25}$ | $c_{17}$ to $c_{23}$ | $c_{10}$ to $c_{14}$ | $d_8$ to $d_{20}$ | $d_{65}$ to $d_{69}$ | $d_{50}$ to $d_{56}$ | $d_{19}$ to $d_{27}$ |
| $c_{16}$ to $c_{22}$ | $c_{20}$ to $c_{28}$ | $c_{20}$ to $c_{24}$ | $c_4$ to $c_{10}$ | $d_{41}$ to $d_{49}$ | $d_{60}$ to $d_{68}$ | $d_{30}$ to $d_{38}$ | $d_8$ to $d_{20}$ |
| - | - | - | - | $d_{12}$ to $d_{20}$ | $d_{68}$ to $d_{72}$ | - | - |
| - | - | - | - | - | $d_{68}$ to $d_{74}$ | - | - |
| - | - | - | - | - | $d_{62}$ to $d_{74}$ | - | - |

**(B) High temperature susceptibility series**

### (i) Estimation of $K_c$

The coefficients chosen are as follows

| Range of coefficients | | | | | | | |
|---|---|---|---|---|---|---|---|
| Series coefficients using $\omega$ | | | | Series coefficients using $\omega_1$ | | | |
| b.c.c | f.c.c | s.c | d. | b.c.c | f.c.c | s.c | d. |
| $g_{15}$ to $g_{22}$ | $g_6$ to $g_{13}$ | $g_{15}$ to $g_{22}$ | $g_{15}$ to $g_{20}$ | $h_{17}$ to $h_{24}$ | $h_7$ to $h_{15}$ | $h_{19}$ to $h_{26}$ | $h_{14}$ to $h_{21}$ |
| $g_3$ to $g_{12}$ | $g_0$ to $g_{11}$ | $g_{11}$ to $g_{20}$ | $g_{13}$ to $g_{20}$ | $h_{12}$ to $h_{17}$ | $h_3$ to $h_{12}$ | $h_7$ to $h_{14}$ | $h_{15}$ to $h_{22}$ |
| $g_1$ to $g_{12}$ | $g_5$ to $g_{14}$ | $g_{11}$ to $g_{22}$ | $g_{12}$ to $g_{17}$ | $h_{13}$ to $h_{20}$ | $h_1$ to $h_{14}$ | $h_{14}$ to $h_{21}$ | $h_6$ to $h_{13}$ |
| $g_1$ to $g_8$ | $g_3$ to $g_{14}$ | $g_8$ to $g_{21}$ | $g_{11}$ to $g_{16}$ | - | - | $h_{17}$ to $h_{24}$ | $h_{10}$ to $h_{17}$ |

### (ii) Estimation of $\gamma$

The coefficients chosen are given below

| Range of coefficients | | | | | | | |
|---|---|---|---|---|---|---|---|
| Series coefficients using $\omega$ | | | | Series coefficients using $\omega_1$ | | | |
| b.c.c | f.c.c | s.c | d. | b.c.c | f.c.c | s.c | d. |
| $l_{18}$ to $l_{24}$ | $l_6$ to $l_{12}$ | $l_{10}$ to $l_{16}$ | $l_1$ to $l_7$ | $m_5$ to $m_{11}$ | $m_7$ to $m_{13}$ | $m_{17}$ to $m_{23}$ | $m_{20}$ to $m_{26}$ |
| $l_8$ to $l_{14}$ | $l_8$ to $l_{14}$ | $l_{12}$ to $l_{18}$ | $l_8$ to $l_{14}$ | $m_{10}$ to $m_{16}$ | $m_2$ to $m_8$ | $m_{12}$ to $m_{18}$ | $m_{18}$ to $m_{24}$ |
| $l_{15}$ to $l_{21}$ | $l_5$ to $l_{11}$ | $l_7$ to $l_{13}$ | $l_5$ to $l_{11}$ | $m_{18}$ to $m_{24}$ | $m_9$ to $m_{15}$ | $m_{16}$ to $m_{22}$ | $m_{16}$ to $m_{22}$ |
| - | $l_7$ to $l_{13}$ | $l_{14}$ to $l_{20}$ | $l_{10}$ to $l_{16}$ | $m_{20}$ to $m_{26}$ | $m_4$ to $m_{10}$ | $m_{22}$ to $m_{28}$ | $m_{13}$ to $m_{19}$ |
| - | - | $l_{18}$ to $l_{24}$ | $l_{15}$ to $l_{21}$ | $m_{15}$ to $m_{21}$ | - | - | - |

**References**

1. R K Pathria and Paul D. Beale (2022) Statistical Mechanics Fourth Edition, Elsevier Academic Press.
2. L Onsager (1944) Phys. Rev., **65** 117.
3. A. L. Talapov and H. W. J. Blöte, (1996) J Phys. **A29** 5727.
4. R. Haggkvist, A. Rosengren, P.H. Lundow, K. Markström, D. Andren and P. Kundrotas, (2007) Adv. Phys. **56** 653.
5. R. Haggkvist, A. Rosengren, D. Andren, P. Kundrotas, P.H. Lundow and K. Markström, (2004) J. Stat Phys. **114** 455.
6. P.H. Lundow, K. Markström and A. Rosengren (2009) Phil.Mag **89** 2009.
7. A M. Ferrenberg and D.P. Landau, (1991) Phys Rev **B44** 5081.
8. M. Hasenbusch, (2010) Phys. Rev. **B 82** 174433.
9. M.Hasenbusch (2016) Phys Rev **E 93** 032140 and references therein.
10. G.A.Baker Jr, (1961) Phys Rev **124** 768.
11. C J Thompson (1972) Mathematical Statistical Mechanics, MacMillan New York.
12. C.Domb in C. Domb and M.S. Green (Ed), (1974) Phase Transitions and Critical Phenomena, Volume 3 Series Expansions for Lattice Models, Academic Press.
13. Gyan Bhanot, Michael Creutz, Ivan Horvath, Jan Lacki and John Weckel, (1993) Phys Rev **E49** 2445.
14. A J Guttmann and I G Enting (1993) J Phys A Math and Gen **26** 807.
15. M F Sykes, D S Gaunt, P D Roberts and J A Wylest (1972) J Phys A: Gen. Phys **5** 640.
16. M F Sykes, J W Essam and D S Gaunt (1965) J Math Phys **6** 283.
17. Başer Tambaş, (2023) Entropy **25** 197.
18. T.Kaya (2022) Chin. J. Phys., **77** 2676
19. T.Kaya (2022) Eur. Phys. J. **13 7**1130.
20. G M. Viswanathan, Marco Aurelio G. Portillo, Ernesto P. Raposo and Marcos G. E. da Luz (2022) Entropy **24** 1665.
21. M V Vismaya and M V Sangaranarayanan (2024) ArXiv 2406.08531.
22. P.Butera, M.Comi and A J Guttmann (2003) Phys Rev **B 67** 054402.
23. D. Shanks, (1955) Nonlinear Journal of Mathematics and Physics, **34** 1.
24. P Wynn, (1956) Math. Tables and other Aids to Computation **10** 91.
25. M.V. Sangaranarayanan and S.K Rangarajan, (1983) Chem Phys Lett **101** 49.
26. M.V. Sangaranarayanan and S.K Rangarajan (1983) Phys Lett **A 96** 339.
27. Gustavo A. Arteca, Francisco M. Fernandez and Eduardo A. Castro (1985) Phys Lett **A 116** 269.
28. A J Guttmann and C J Thompson, (1969) Phys Lett **A 28** 679.
29. A J Guttmann (1989) in Phase Transitions and Critical Phenomena Volume 13, (Ed) C Domb and J L Lebowitz (Academic Press New York).
30. G.A. Baker Jr. and P. Graves-Morris, (1981) in: Encyclopaedia of mathematics, Vol. 13, ed. G.C. Rota (Addison- Wesley, Reading)
31. Wakefield (1951) Proc Cambridge Phil Soc **47** 799.
32. Francisco Sastre, (2021) Physica **A 572** 125881.
33. A Rosengren, (1986) J Phys A Math and Gen **19** 1709.

34. J.Zinn-Justin J (1981), J. Physique **42** 783.
35. D S Gaunt (1982) in Phase Transitions ed M Levy, J C Le Guillou and J Zinn-Justin (New York: Plenum) p 217.
36. J. Adler (1983), J. Phys. A: Math. Gen. **16** 3585.
37. G S Pawley, R H Swendsen, D J Wallace and K G Wilson (1984) Phys. Rev. **B 29** 4030.
38. M N Barber, R B Pearson, D Toussaint and J L Richardson (1983) Preprint NSF-ITP-83-144.
39. R B Pearson, (1984) Phys. Rep. **103** 185.
40. U.Yu (2015) Physica A Stat. Mech. Appl **419** 75.
41. Xiang-Ke Chang,Yi He, Xing-Biao Hu , Jian-Qing Sun and Ernst Joachim Weniger (2020) Numerical Algorithms **83** 593.
42. Mehran Kardar (2007) Statistical Physics of Fields, Cambridge University Press.
43. A.Pellissetto and E.Vicari, (2002), Phys Rep **368** 549.
44. A J Liu and M E Fisher, (1990) J Stat Phys **58** 431.
45. M E Fisher and Jacques H.H. Perk, (2016) Phys Lett **A 380** 1339 and references therein.
46. J.V. Sengers and J.G. Shanks, (2009) J. Stat. Phys, **137** 857.
47. Zaher Salman and Joan Adler, (1998), Int J Mod Phys **C9** 195.
48. I.A. Hadjiagaplou, A. Malakis and S.S. Martinos (2006), Rev.Adv. Mater.Sci **12** 63.
49. N.Ito and M.Suzuki J, Phys Soc Japan 60(1991)1978.
50. A.M.Ferrenberg, J.Xu and D P Landau, (2018) Phys. Rev. **E 97** 043301.